\documentclass[a4paper,11pt]{article}

\usepackage[utf8]{inputenc}

\usepackage[backend=biber,style=numeric-comp,sorting=none]{biblatex}
\newcommand{\TheAuthor}{Simone Romiti}
\newcommand{\TheEmail}{simone.romiti.1994@gmail.com}
\newcommand{\TheTitle}{$\mathrm{SU(N)}$ lattice gauge theories with Physics-Informed Neural Networks}

\usepackage{authblk}
\usepackage{orcidlink}
\usepackage{amsmath}
\usepackage{amssymb}
\usepackage{dsfont}
\usepackage{MnSymbol} 
\usepackage{braket}
\usepackage{graphicx}
\usepackage{xcolor}
\usepackage{tikz}
\usetikzlibrary{positioning, chains, shapes.geometric, fit, shapes, arrows.meta, calc, backgrounds}

\usepackage{algorithm}
\usepackage{algpseudocode}

\numberwithin{equation}{section}

\usepackage{hyperref}
\hypersetup{
    pdftitle={\TheTitle{}},
    pdfauthor={\TheAuthor{}},
    bookmarks,
    colorlinks,
    linkcolor=blue,
    citecolor=blue,
    menucolor=black,
    urlcolor=blue,
    filecolor=blue
}

\begin{document}

\author{
    \TheAuthor{}~\orcidlink{0000-0002-6509-447X}~\thanks{Email:
        \href{mailto:\TheEmail}{\TheEmail{}}}
}
\affil{Institute for Theoretical Physics, Albert Einstein Center for Fundamental Physics, University of Bern, CH-3012 Bern, Switzerland}
\title{\TheTitle{}}
\date{\today{}}

\maketitle

\begin{abstract}
    We present an application of Physics-Informed Neural Networks (PINNs) to the study of $\mathrm{SU}(N_c)$ lattice gauge theories.
Our method enables the learning of eigenfunctions and eigenvalues at arbitrary gauge couplings, smoothly moving from the analytically known strong-coupling regime towards weaker couplings.
By encoding the Schrödinger equation and the symmetries of the eigenstates directly into the loss function, the network performs an unsupervised exploration of the spectrum.
We validate the approach on the single-plaquette $\mathrm{U}(1)$ and $\mathrm{SU}(2)$ pure-gauge theories, showing that the PINNs successfully reproduce the hierarchy of energy levels and their corresponding wavefunctions.
\end{abstract}



\section{Introduction}
\label{sec:introduction}

The lattice formulation allows to study strongly-interacting Quantum Field Theories from the first principles of the theory~\cite{gattringer2009quantum,degrand2006lattice}. 
In this formalism, Monte Carlo (MC) techniques are able to approximate the path integral,
so that one can extract physical observables from correlation functions~\cite{FlavourLatticeAveragingGroupFLAG:2024oxs}.
However, despite their successes, MC techniques suffer from fundamental limitations, such as the impossibility to simulate real-time dynamics, the sign problem~\cite{gattringer2016approaches}, topological freezing~\cite{Luscher:2011kk,Mazur:2021zgi} and critical slowing down~\cite{Schaefer:2010hu,Cossu:2017eys}. 

The above limitations are absent if we resort to the Hamiltonian formulation, which works directly at the level of the Schr\"odinger equations~\cite{sakurai_napolitano_2017}.
In recent years this approach has seen a revived interest through Quantum Computing (QC)~\cite{Funcke:2023jbq} and Tensor Networks (TNs)~\cite{Magnifico:2024eiy},
which work by truncating the infinite-dimensional Hilbert space end extrapolating the result to the infinite-truncation limit. 
While these techniques remain promising to mitigate Hilbert space size scaling cost, several present applications are still limited to a moderate number of lattice degrees of freedom.

In this paper we propose and explore a novel approach based on Physics-Informed Neural Networks (PINNs)~\cite{cuomo2022scientific}, applied to the solution of the eigenvalue equation for $\mathrm{SU}(N_c)$ lattice gauge theories.
We use Machine Learning to build models that represent the wavefunctions of the lattice Hamiltonian, and in the learning process we also learn their corresponding eigenvalues.
The result is a prediction of the ladder of energies and eigenstates of the theory, 
for ranges of the coupling where their analytic expression is not known.
In this work we show that it is possible to achieve this goal by using \textit{adiabatic PINNs}: we start from the strong-coupling regime, where we are able to write the eigenstates analytically, and move progressively to the weak-coupling regime~(see Sec.~\ref{sec:PINNsStrategy} for more details).

For the training of PINNs we sample points in the function domain and penalize, through the loss function, the violation of specific integral and differential constraints. 
In our case these are the eigenvalue equation, the symmetries of the wavefunction and its normalization.
This feature allows to solve the time-independent Schr\"odinger equation at non-trivial couplings, without the need of a validation set from the target eigenfunction~\cite{moseley2020solving}.
Moreover, PINNs do not impose an explicit truncation of the Hilbert space. 
The accuracy of the result is dictated by the quality of the sampled configurations and choice of the loss function.

In this work we develop the general method for $\mathrm{SU}(N_c)$ gauge theories.
In order to demonstrate and benchmark the approach, we reproduce the known analytic results for pure-gauge $\mathrm{U}(1)$ and $\mathrm{SU}(2)$ lattice gauge theories, in the limit of a single plaquette system~\cite{Ligterink:2000ug,Bender:2020jgr}.

The paper is structured as follows.
In Sec.~\ref{sec:theory} we recall some theoretical background. Sec.~\ref{sec:PINNsStrategy} describes our adiabatic PINN method and in Sec.~\ref{sec:OnePlaquetteSystem} we show our results for the single-plaquette system. Finally, in Sec.~\ref{sec:Conclusions} we draw our conclusions and discuss future developments.
In order to improve the readability of the main sections, we have collected some more technical details in the Appendices. 
In App.~\ref{sec:HamiltonianLGT} we recall the formulation of the $\mathrm{SU}(N_c)$ lattice Hamiltonian, and in App.~\ref{sec:GaugeFixing} we describe the gauge-fixing procedure.

\section{Theoretical background}
\label{sec:theory}

In quantum mechanics~\cite{sakurai_napolitano_2017}, solving a theory requires to fully identify the physical states by their symmetries and solving the time-independent (i.e.~eigenvalue) Schr\"odinger equation:
\begin{equation}
    H \ket{\vec{\alpha}, E} = E \ket{\vec{\alpha}, E}
    \, .
\end{equation}
Here $H$ is the Hamiltonian, $E$ is the energy and $\vec{\alpha}$ is the set
of all the other quantum numbers dictated by the eigenvalues of the symmetries
operators. 
At this point the time-dependent Schr\"odinger equation dictates how the states evolve in time.

Except for a few examples~\cite{Sasaki:2014tka}, solving analytically the eigenvalue equation is in general not possible, requiring the use of numerical tools. 
An example is the case of strongly-interacting $\mathrm{SU}(N_c)$ gauge theories~\cite{peskin2018introduction}. 
On the lattice~\cite{RevModPhys.51.659,PhysRevD.11.395}, in the wavefunction formalism, the eigenvalue equation becomes a Partial Differential Equation (PDE) coupling neighboring degrees of freedom.
For instance, in a pure-gauge theory it reads~(c.f.~Eq.~\eqref{eq:HamiltonianHBplusHE})~\footnote{
    We work in natural units, $\hbar = c = 1$, and measure all quantities in units of the lattice spacing.
    The sign of the coupling $g$ is immaterial; we adopt the convention $g > 0$ throughout.
}:
\begin{equation}
    \label{eq:SchroedingerEqSUN}
    \left[
    \frac{g^2}{2} \sum_{\vec{x}} \sum_{\mu=1}^{d-1} \sum_{a=1}^{N_g}
    (L_a)_{\mu}^2(\vec{x})
    - \frac{2}{g^2} \sum_{\vec{x}} \sum_{\mu=1, \nu < \mu}^{d-1}
    \text{Re} \text{Tr} [\mathcal{U}_{\mu\nu}(\vec{x})]
    \right] 
    \, \psi(\mathcal{U}) = E \, \psi(\mathcal{U})
    \, ,
\end{equation}
where ${\psi(\mathcal{U}) = \braket{\mathcal{U}|\psi}}$ is the wavefunction of the whole configuration of gauge links.
The individual links ${\mathcal{U}_\mu(\vec{x})}$ are elements of the gauge group $\mathrm{SU}(N_c)$, and can be parametrized in terms of $N_g$ angles $(\theta^a)_\mu(\vec{x})$, where $N_g$ is the number of generators of the Lie algebra~(see Sec.~\ref{sec:HamiltonianLGT}).
Since, for each $\vec{x}$ and $\mu$, the $L_a$ are differential operators in the angles $\theta^a$, Eq.~\eqref{eq:SchroedingerEqSUN} is in fact a PDE in the variables $(\theta^a)_\mu(\vec{x})$.

The eigenvalue equation does not fully determine the properties of the state.
First, the eigenvalue may be degenerate, requiring to specify the other quantum numbers in order to fully determine it.
Second, the wavefunction can be multiplied by an arbitrary constant complex number (phase and normalization).
The choice of the phase can be fixed by,~e.g., imposing that $\psi$ should be real at given configuration, while the normalization is fixed by imposing the total probability to be $1$~\cite{sakurai_napolitano_2017,gattringer2009quantum}:
\begin{equation}
    \int \mathcal{D}\mathcal{U} \, |\psi(\mathcal{U})|^2 = 1
    \, ,
\end{equation}
where $\mathcal{D}\mathcal{U}$ is the Haar measure of the gauge group.
Finally, the physical states must be gauge-invariant,~i.e.:
\begin{equation}
    G_a(\vec{x}) \, \psi(\mathcal{U}) = 0 , \quad \forall \, a, \vec{x}
    \, .
\end{equation}
The $G_a(\vec{x})$ are are also differential operators in the angles $(\theta^a)_\mu(\vec{x})$, that represent the generators of gauge transformations (see Sec.~\ref{sec:HamiltonianLGT}).

\section{Strategy for training the Neural Network}
\label{sec:PINNsStrategy}
In this section we describe our general strategy to solve the time-independent
Schr\"odinger equation with PINNs. 
We discuss how to find numerically the dependence of a set of eigenstates (and eigenvalues) as a function of the gauge coupling, moving away \textit{adiabatically} from the strong-coupling limit. 

\subsection{Adiabatic training and existence of the solution.}
\label{sec:AdiabaticTrainingStrategy}

Training a network at an arbitrary coupling in order to ``just solve'' the eigenvalue equation is an ill-posed numerical problem: in general there exist multiple solutions with the same quantum numbers except for the energy. 
This does not simply make the convergence to the solution very hard, but even if successful it does not inform about where in the spectrum hierarchy one landed.
For these reasons, we propose an algorithm defined in terms of the adiabatic approximation theorem of quantum mechanics~\cite{sakurai_napolitano_2017}.
According to the theorem, a physical state remains in its instantaneous
eigenstate if the Hamiltonian $H$ changes slowly over time and a energy gap with the other states is preserved.
On a finite volume, the latter condition is fulfilled because the spectrum is discrete for $\mathrm{SU}(N_c)$ theories on a torus~\cite{Luscher:1982ma}.
We apply the concept of the adiabatic evolution to an Hamiltonian whose coupling changes over the training time, and interpret the training process from one coupling to the other as a quantum time evolution with a time-dependent Hamiltonian.
In this sense, the algorithm presented in this work predicts the coupling dependences of eigenfunctions and energies of the system, using the strong-coupling limit as a starting point and flowing to larger values of $1/g$.

\begin{figure}[H]
    \begin{center}        
        \includegraphics[width=0.9\textwidth]{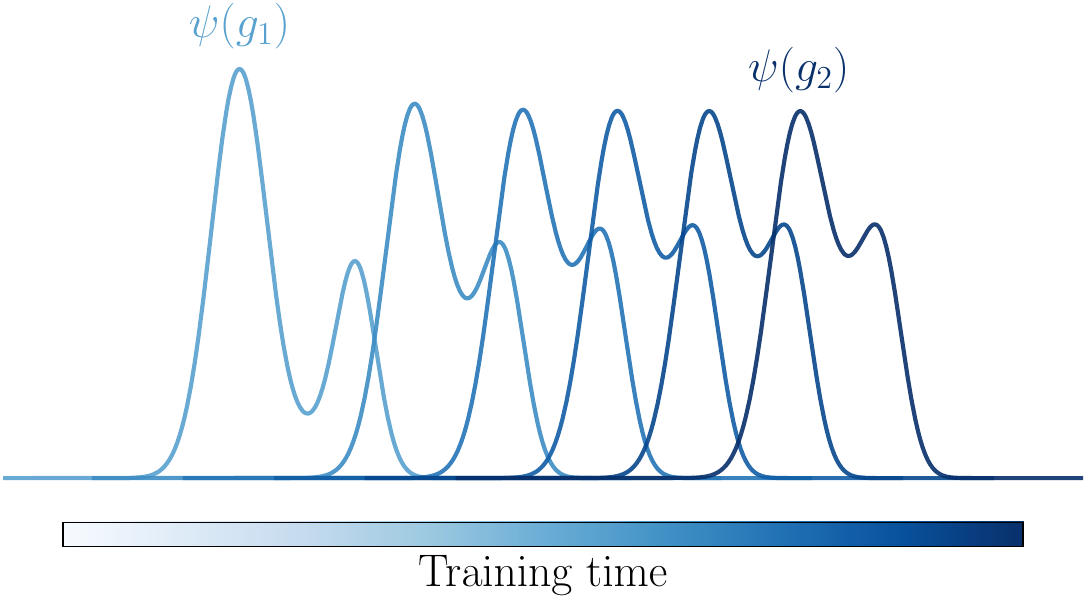}
    \end{center}
    \caption{
        Pictorial representation of the ideal adiabatic learning described in the text.
        The normalization of the state and the eigenvalue equation are imposed throughout, while we evolve the coupling smoothly from $g_1$ to $g_2$ (from left to right). 
        The eigenfunction $\psi(g_1)$ remains in its instantaneous eigenstate, with final form given by $\psi(g_2)$.
    }
    \label{fig:AdiabaticTraining}
\end{figure}

The adiabatic theorem also guarantees the existence of the solution according to this procedure.
In fact, let us assume that we have a perfect model that describes our normalized eigenstate
of $H(g_1)$ at some coupling $g_1$. 
We now refine it to become an eigenstate of $H(g_2)$ and by enforcing the normalization condition exactly while training. 
Since the norm does not change, by definition, the latter condition means that we are
evolving the state with a unitary operator. 
If our model converges to an eigenstate of $H(g_2)$, we have performed a time-evolution of the eigenstate from the couplings $g_1$ to $g_2$. 
The adiabatic theorem ensures that if we have done it sufficiently slowly we have not moved from the instantaneous eigenstate. 
In Fig.~\ref{fig:AdiabaticTraining} we show a pictorial representation of this process.

We remark that in the ${1}/{g} \to 0$ limit the wavefunction of the electric vacuum $\ket{0}_\text{E}$ is known analytically.
Moreover, by acting with gauge-invariant loop operators on $\ket{0}_\text{E}$ we obtain the expression of the electric eigenstates and an (overcomplete) basis for the Hilbert space of physical states~\cite{PhysRevD.11.395,Mathur:2005fb,Mathur:2007nu}.
Therefore, for each strong-coupling eigenstate, we train a network on the analytic result and progressively fine-tune it at subsequent couplings.

Summarizing, our method consists in considering the $n$-th eigenvalue at $1/g \to 0$, and train the neural network on a set of sampled configurations on the analytic values. 
At this point, we load the model and train with a slightly different new coupling.
During the training we learn the new eigenfunction and energy (treated as a training parameter)  by imposing only physical constraints. 
A schematic representation of the protocol is given by the Algorithm~\ref{algo:AdiabaticProtocol}.
\begin{algorithm}[H]
    \caption{Pseudo-adiabatic protocol for PINN training at coupling $g^{-1}$}
    \label{algo:AdiabaticProtocol}
    \begin{algorithmic}[1]
        \State \textbf{Input:} Coupling $g^{-1}$, small step size $\Delta g^{-1}$
        \If{$g^{-1} = 0$}
        \State Initialize new model. Energy known exactly.
        \State Train with MSE loss on sampling points
        \Else
        \State Load model from previous coupling. $g_\text{old}^{-1} = (g^{-1} - \Delta g^{-1})$
        \State Energy not known, treated as training parameter.
        \State Train with composite PINN loss function. 
        \EndIf
        \State \textbf{Output:} Trained eigenfunction $\psi$ and energy $E(g^{-1})$.
    \end{algorithmic}
\end{algorithm}

The idea of using PINNs comes from the fact that it is a form of unsupervised learning for a model (and potentially external parameters) without prior knowledge on the values of the target function.
One of their advantages is that Neural Networks offer an easy way of implementing auto-differentiation, so that the derivatives of the model with respect to the inputs can be computed exactly.
In this way one can impose differential and boundary conditions that completely identify the solution of a PDE.
The element of innovation here is that we exploit the analytic results in a given coupling regime ($1/g \to 0$), and use them as a seed to flow adiabatically to other values of the coupling.

We remark that the above adiabatic argument is theoretical: our initial model will
hardly be perfect, we do not know \textit{a priori} how slowly we should change
the coupling, and numerically we can not enforce the normalization condition
exactly. 
The latter limitation stems not simply from machine precision, but
also from the fact that since at each iteration we do not know the eigenfunction
at the new coupling, we inevitably commit sampling errors when estimating the
normalization integral. 
However, up to these numerical approximation problems, we stress that the adiabatic theorem guarantees the validity of the protocol.
Moreover, we can always check \textit{a posteriori} the consistency of our results by, e.g, changing the coupling at lower rates, activating smoothly different contributions to the Hamiltonian, or even running the algorithm backwards (from weaker towards stronger couplings), and verify that we get the same result at intermediate values.

\subsection{Choice of the loss function}
\label{sec:LossFunctionChoice}

In this section we describe our choice for the loss function.
We limit ourselves to pure-gauge theories, deferring the study of theories with fermions to a future work.
The latter generalization does not add further differential operators, but requires to simply encode the dependency on the fermions configurations as well as the action of the Hamiltonian on the wavefunction.

In Sec.~\ref{sec:AdiabaticTrainingStrategy} we have described how in principle one can take an eigenstate $\ket{\phi_n}$ in the strong-coupling limit and predict the coupling flow of its wavefunction and energy.
We realize this by taking the analytic values of $\phi_n$ and training the model on them.
For this we use the Mean-Square-Error (MSE) loss function:
\begin{equation}
    \mathcal{L}^{(1/g \to 0)}
    =
    \frac{1}{N_{\text{pt}}} 
    \sum_{i=1}^{N_\text{pt}} |\psi(\mathcal{U}_i) - \phi_n(\mathcal{U}_i)|^2
    \, ,
\end{equation}
where $N_\text{pt}$ is the number of points sampled in configuration space, and the energy is fixed to its exact value during the training.
This number can be changed at will for different couplings, while the strong-coupling provides a rough estimate for the necessary network size.

At progressively weaker couplings, we first load the model at the previous coupling, 
and enforce the time-independent Schr\"odinger equation both on the wavefunction $\psi$ and energy $E$ (which is now a trainable parameter).
We use the following loss function:
\begin{equation}
    \mathcal{L}^{(1/g \, \neq \, 0)} =
    \omega_\text{PDE} \mathcal{L}_\text{PDE} +
    \omega_\text{norm} \mathcal{L}_\text{norm} +
    \omega_\text{symm} \mathcal{L}_\text{symm}
    \, ,
\end{equation}
where $\omega_\text{PDE}$, $\omega_\text{norm}$, $\omega_\text{symm}$ are meta-parameters of the training that act as weights for the gradients of the loss function contributions.
The individual terms are:
\begin{align}
    \label{eq:LossPDE}
    \mathcal{L}_\text{PDE}  & =
    \frac{1}{N_\text{pt}} \sum_{i=1}^{N_\text{pt}}
    |(H-E) \psi(\mathcal{U}_i)|^2
    \, ,
    \\
    \label{eq:LossNorm}
    \mathcal{L}_\text{norm} & =
    \left( 1 - \sqrt{\int d \mathcal{U} \, |\psi(\mathcal{U})|^2}
    \right)^2
    \, ,
\end{align}
where $\mathcal{U}_i$ is the $i$-th sampled configuration of links,
$\mathcal{L}_\text{symm}$ is a term that enforces the specific symmetries of the eigenstate in question and gauge invariance through Gauss' law. 
The latter is imposed through a term $\mathcal{L}_\text{Gauss}$ in the loss function:
\begin{equation}
    \mathcal{L}_\text{Gauss}  =
    \frac{1}{N_\text{pt}} \sum_{i=1}^{N_\text{pt}}
    \sum_{\vec{x}}
    \sum_a
    |G_a(\vec{x}) \psi(\mathcal{U}_i)|^2
\end{equation}
%


In practice, for the eigenfunctions studied in this work, we found that it was sufficient to impose ${\omega_\text{PDE}=\omega_\text{norm}=\omega_\text{symm}=1}$.

The norm entering in the loss of Eq.~\eqref{eq:LossNorm} is obtained at each training step as a Monte Carlo approximation over $N_\text{pt}$ configurations:
\begin{equation}
    \int d \mathcal{U} \, |\psi(\mathcal{U})|^2 
    \approx
    \Omega \frac{1}{N_\text{pt}} \sum_{i=1}^{N_\text{pt}} |\psi(\mathcal{U}_i)|^2
    \, ,
\end{equation}
where $\Omega$ is the total Haar measure of all the links~\cite{gattringer2009quantum}. 
We remark that this introduces a sampling error, as we do not know yet the analytic form of the wavefunction and we necessarily have to sample from the one at the previous coupling.
Moreover, even if we knew $\psi$ exactly, the sampling could still remain a challenging task due to multi-modalities of the $|\psi(\mathcal{U})|^2$ distribution. 
In the numerical calculations of this work it was sufficient to always finely sample the phase space uniformly. 
For future investigations, it would be interesting to consider Nested Sampling~\cite{Ashton:2022grj}, which is well-suited for exploring multi-modal distributions. Work on applying this method to gauge theories is presently underway~\cite{NestedSamplingKanwarRomitiWenger2025}.



\subsection{Network architecture and learning process}

In this work we use Fully-Connected Neural Networks (FCNNs) with \texttt{tanh} activation function, as model templates for the eigenfunctions. 
This choice is led by the universal approximation
theorems of neural networks~\cite{augustine2024survey} and the fact that for the physical
systems and regimes studied in this work, these networks proved to be complex
enough to capture the physical features of the eigenstates. 
While the same choice was successful in other applications of PINNs~\cite{jin2020unsupervised,jin2022physics,Holliday:2023ugi,bonder2025pinns,rowan2025solving}, we recognize
that it would be interesting to understand in a future work how different network architectures
affect the efficiency.

\begin{figure}[H]
    \begin{center}
        \begin{tikzpicture}[
    >=LaTeX, 
    node/.style={ 
        circle,
        minimum width=2.25em,
        draw,
        fill=gray!10,
        thick
    },
    cell/.style={
        rectangle,
        rounded corners=2mm,
        minimum height=2.5em,
        minimum width=2.5em,
        draw,
        thick
    }, 
    arrow/.style={
        -latex,
        thick
    },
    backprop/.style={ 
        arrow,
        dashed,
        gray
    }
]
    	
    \foreach \x in {1,...,2}
        \draw node at (0, -\x*1.25 - 0.625) [node] (first_\x) {$\mathcal{U}_\x$};
    \draw node at (0, -5*1.25 + 0.625) [node] (first_n) {$\mathcal{U}_{N_\ell}$};
    \path (first_2) -- (first_n) node[pos=0.38, scale=2] {\vdots};
    	
    \foreach \x in {1,...,3}
        \node at (2.5, -\x*1.25) [node] (second_\x){};
    \draw node at (2.5, -5*1.25) [node] (second_m) {};
    \path (second_3) -- (second_m) node[pos=0.38, scale=2] {\vdots};

    \foreach \x in {1,...,3}
        \node at (5, -\x*1.25) [node] (third_\x){};
    \draw node at (5, -5*1.25) [node] (third_m) {};
    \path (third_3) -- (third_m) node[pos=0.38, scale=2] {\vdots};
    	
    \node at (7.5, -3*1.25) [node] (psi) {$\psi$};
    		
    \foreach \i in {1,...,2}
        \foreach \j in {1,...,3}
            \draw [arrow] (first_\i) to (second_\j);
    \foreach \i in {1,...,2}
        \draw [arrow] (first_\i) to (second_m);
    \foreach \i in {1,...,3}
        \draw [arrow] (first_n) to (second_\i);
    \draw [arrow] (first_n) to (second_m);

    \foreach \i in {1,...,3}
        \foreach \j in {1,...,3}
            \draw [arrow] (second_\i) to (third_\j);
    \foreach \i in {1,...,3}
        \draw [arrow] (second_\i) to (third_m);
    \foreach \i in {1,...,3}
        \draw [arrow] (second_m) to (third_\i);
    \draw [arrow] (second_m) to (third_m);
    	
    \foreach \i in {1,...,3}
        \draw [arrow] (third_\i) to (psi);
    \draw [arrow] (third_m) to (psi);
    	
    \node[above=0.25cm of first_1] (xlabel) {$\mathcal{U}$};
    \node[above=0.25cm of second_1] (h1label) {$\mathbf{h}^{(1)}$};
    \node[above=0.25cm of third_1] (h2label) {$\mathbf{h}^{(2)}$};
    \node[above=0.25cm of psi] (ylabel) {$\psi(\mathcal{U})$};

    \node[cell, dashed, fit=(first_1) (first_n)] (xbox) {};
    \node[cell, dashed, fit=(second_1) (second_m)] (xbox) {};
    \node[cell, dashed, fit=(third_1) (third_m)] (xbox) {};
    \node[cell, dashed, fit=(psi)] (xbox) {};
\end{tikzpicture}
    \end{center}
    \caption{
        Illustrative representation of the Fully-Connected-Neural-Network used in this work in train the eigenfunctions.
        The inputs are the gauge links of the lattice (in the form of the angles needed to parametrize them), which are passed to the hidden layers $h^{(1)}$, $h^{(2)}$ to output the wavefunction $\psi$. Each layer uses a \texttt{tanh} activation function.}
    \label{fig:FCNsketch}
\end{figure}
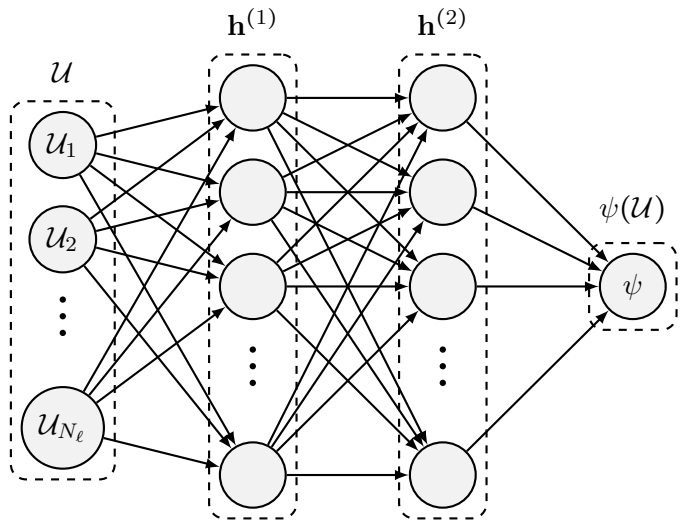

Our networks take as inputs the angles parametrizing the links of the lattice,
pass them to 2 hidden layers made of 512 neurons each throughout, and return the wavefunction value (see Fig.~\ref{fig:FCNsketch}).

In order to avoid overfitting, we have included weight decay~\cite{NIPS1988_1c9ac015} using the \texttt{AdamW} algorithm~\cite{loshchilov2017decoupled} for minimization.
The above choices were sufficient for our proof-of-principle application to demonstrate the method.
We used the strong-coupling limit of the theory ($1/g \to 0$) to get an estimate of the above network size needed to capture the complexity of the wavefunctions.
We used ${N_\text{pt}=10^4}$ points throughout, obtained from a uniform sampling in configuration space. 
At intermediate couplings, we have checked that increasing the network size and number of points was keeping the results stable.

We conclude this section by reporting our strategy for the choice of the learning rate $\epsilon_r$, responsible for the convergence speed towards the minimum of the loss function through gradient descent.
If $\epsilon_r$ is too large, we make too coarse steps and we may never converge, while if it is too small we may spend too much time to find the solution and risk of falling in a local minimum.
The tradeoff we adopted for our trainings in this work was to fix an initial learning rate, and use a ``reduce on plateau'' scheduler for its update~\cite{paszke2019pytorch}~\footnote{While this was sufficient for our study, for the future we plan to consider other unsupervised learning rate schedulers such as \texttt{Twin}~\cite{brigato2024tune}.}.
In practice, we find empirically the initial value of $\epsilon_r$ and of the \texttt{patience}, so that during the training $\epsilon_r$ is automatically reduced when the optimizer is not reducing the loss function within a \texttt{patience} window.

During the network training we used the following stopping criterion.
Since the energy $E$ of the state is a trained parameter, we also construct during the training the following estimator:
\begin{equation}
    \label{eq:EnergyEstimator}
    E_{\text{est.}} = \frac{\int d\mathcal{U} \psi^{*}(\mathcal{U}) [H \psi(\mathcal{U})]}{\int d \mathcal{U} \, |\psi(\mathcal{U})|^2}
    \, ,
\end{equation}
which is the definition of the energy as expectation value.
The ratio ${r=\frac{(E-E_{\text{est.}})}{E}}$ gives a measure of the stability of the training.
For the specific applications of this work it was sufficient to end the training either when $\epsilon_r$ went below $10^{-8}$, or when reaching $10^5$ epochs. In this way, we always obtain $r < 10^{-3}$ at the end of the training.

\section{Results for the single-plaquette system}
\label{sec:OnePlaquetteSystem}

In this section we compare our findings with known analytical results for pure-gauge $\mathrm{U}(1)$ and $\mathrm{SU}(2)$ theories, in the limit of a single-plaquette system. 
The latter are obtained in a gauge-fixed Hamiltonian. 
We defer to Sec.~\ref{sec:GaugeFixing} for a recap on gauge fixing and the corresponding construction of the Hamiltonian.
For a single plaquette, fixing the gauge simplifies the training of the PINNs, as the time-independent Schr\"odinger equation reduces to an ordinary differential equation.
An \texttt{Python} implementation is publicly available at the following URL: 
\\
\href{https://github.com/simone-romiti/adiabatic_PINNs-suN}{https://github.com/simone-romiti/adiabatic\_PINNs-suN}.

Before presenting the specific case of $\mathrm{U}(1)$ and $\mathrm{SU}(2)$ we recall the following properties.
The wavefunction has the form
${\psi(\mathcal{U}_1,\ldots,\mathcal{U}_{N_\ell})}$, where $\mathcal{U}_k$ are the gauge links of the lattice. 
In the strong-coupling
limit, ${1/g \to 0}$, the (electric) ground state $\psi_0$ is the identity
function, and the Hilbert space of gauge invariant wavefunctions is spanned by traces of Wilson loops $\mathcal{L}$ applied to it~\cite{PhysRevD.11.395,Mathur:2005fb,Mathur:2007nu}:
${\operatorname{Tr}[\mathcal{L}] \cdot \psi_0}$.
For a single-plaquette system, in this limit the eigenfunctions are the result of the repeated
application of the only available Wilson loop (the plaquette operator) $n$ times to the electric vacuum.

\subsection{$\mathrm{U}(1)$ gauge theory}
\label{sec:UoneSinglePlaquetteResults}
For $\mathrm{U}(1)$ we have only one generator in the Lie algebra, and each of
the $N_\ell$ links is parametrized by a single angle. 
Therefore, a link and its canonical momentum admit the following representation:
\begin{align}
     &
    \mathcal{U} = e^{i \phi} \, , \quad  \phi \in [0, 2\pi]
    \, ,
    \\
     &
    L = -R = i\frac{\partial}{\partial \phi}
    \, .
\end{align}
The Hamiltonian reads:
\begin{equation}
    H = - \frac{g^2}{2} \sum_{k=1}^{N_\ell} \frac{\partial^2}{\partial \phi_k^2}
    - \frac{2}{g^2} \sum_{\Box} \cos{\phi_\Box}
    \, ,
\end{equation}
where $\phi_\Box$ is the total phase accumulated by multiplying the links
around a plaquette $\Box$. 
When we restrict to a single plaquette and fix the gauge of the maximal tree,
we get the following
Hamiltonian~(c.f.~Eq.\eqref{eq:HamiltonianMaxTreeSinglePlaquette}):
\begin{equation}
    H = - 2 g^2 \frac{d^2}{d \omega^2} - \frac{2}{g^2}
    \cos{\omega}
    \, ,
\end{equation}
where $\omega$ is the angle that parametrizes the only active link.
In this way, the eigenvalue equation ${H \psi = E \psi}$ in the variable
$z=\omega/2$ becomes the Mathieu
equation~\cite{Ligterink:2000ug,Bender:2020jgr}:
\begin{equation}
    \label{eq:MathieuEquationUone}
    \left[ \frac{d^2}{d z^2} + \alpha - 2 q \cos{(2z)} \right] \psi(e^{i2z}) = 0
    \, ,
\end{equation}
where ${\alpha=2E/g^2}$ and ${q=-2/g^4}$. From a purely mathematical viewpoint,
the solutions are the Mathieu functions $\text{ce}_n(z,q)$ and
$\text{se}_n(z,q)$, together with their pseudo-periodic partners from Floquet
theorem (see e.g. Refs.~\cite{Muller-Kirsten:2012wla,abramowitz1965handbook,Robson:1980nt,whittaker2020course}).
Of these solutions however, physically we can consider only the $\pi$-periodic
ones in $z$, as $\psi$ has this period by construction.

In the strong-coupling limit $q \to 0$, and the eigenfunctions are the complex
exponentials ${\psi_n=e^{i n \omega}}$ with eigenvalues ${E_n = 2 g^2 n^2}$. 
As anticipated before, we observe that these are indeed the result of the (untraced) plaquette operator applied to the electric vacuum.

\begin{figure}[H]
    \begin{center}
        \includegraphics[width=\textwidth]{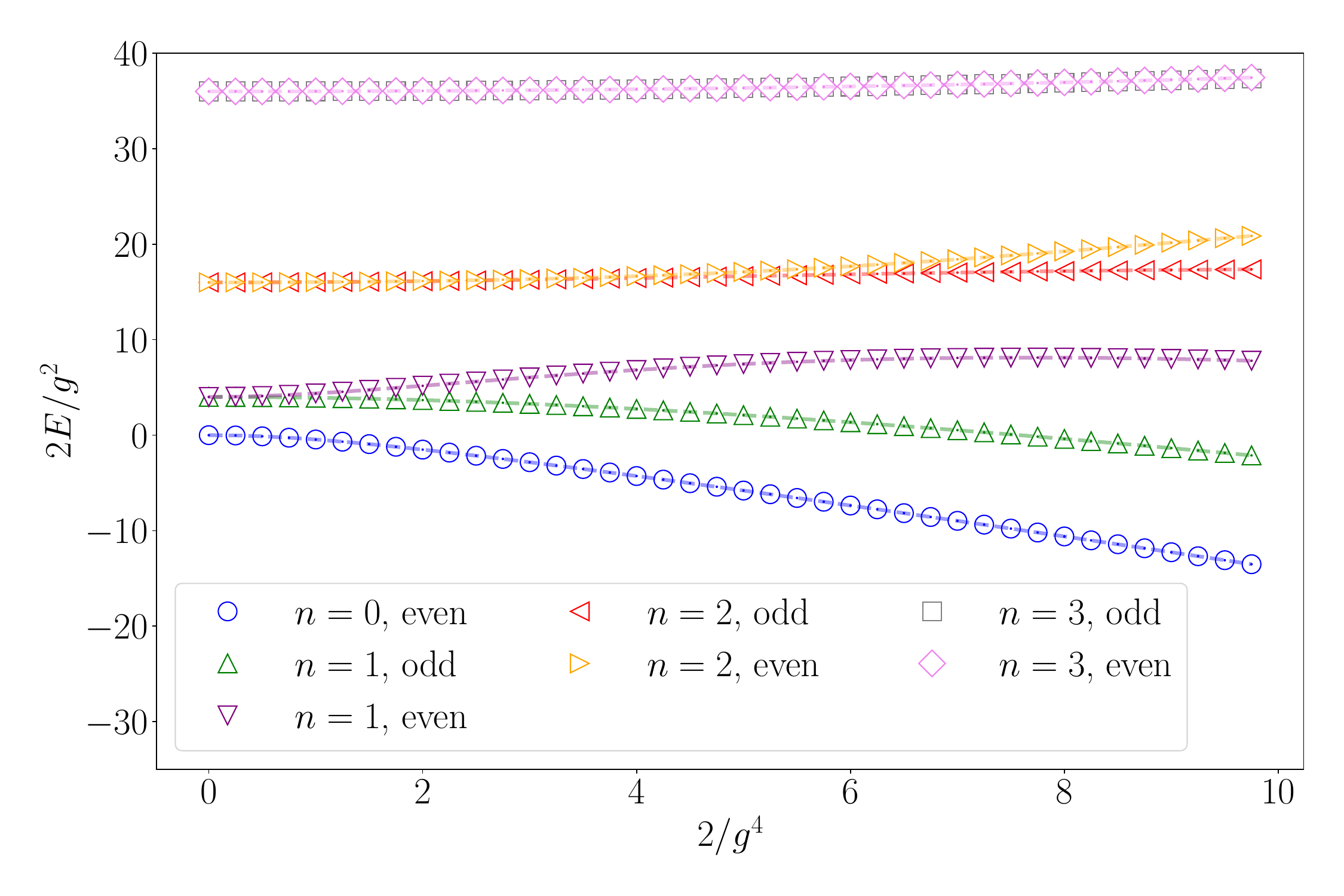}
    \end{center}
    \caption{
        First energy levels for the even and odd eigenfunctions of a $\mathrm{U}(1)$ single-plaquette system.
        The discrete points represent our prediction obtained with a PINN, while the dashed lines correspond to the exact analytic values.}
    \label{fig:U1Energies}
\end{figure}

\begin{figure}[H]
    \includegraphics[width=0.5\textwidth]{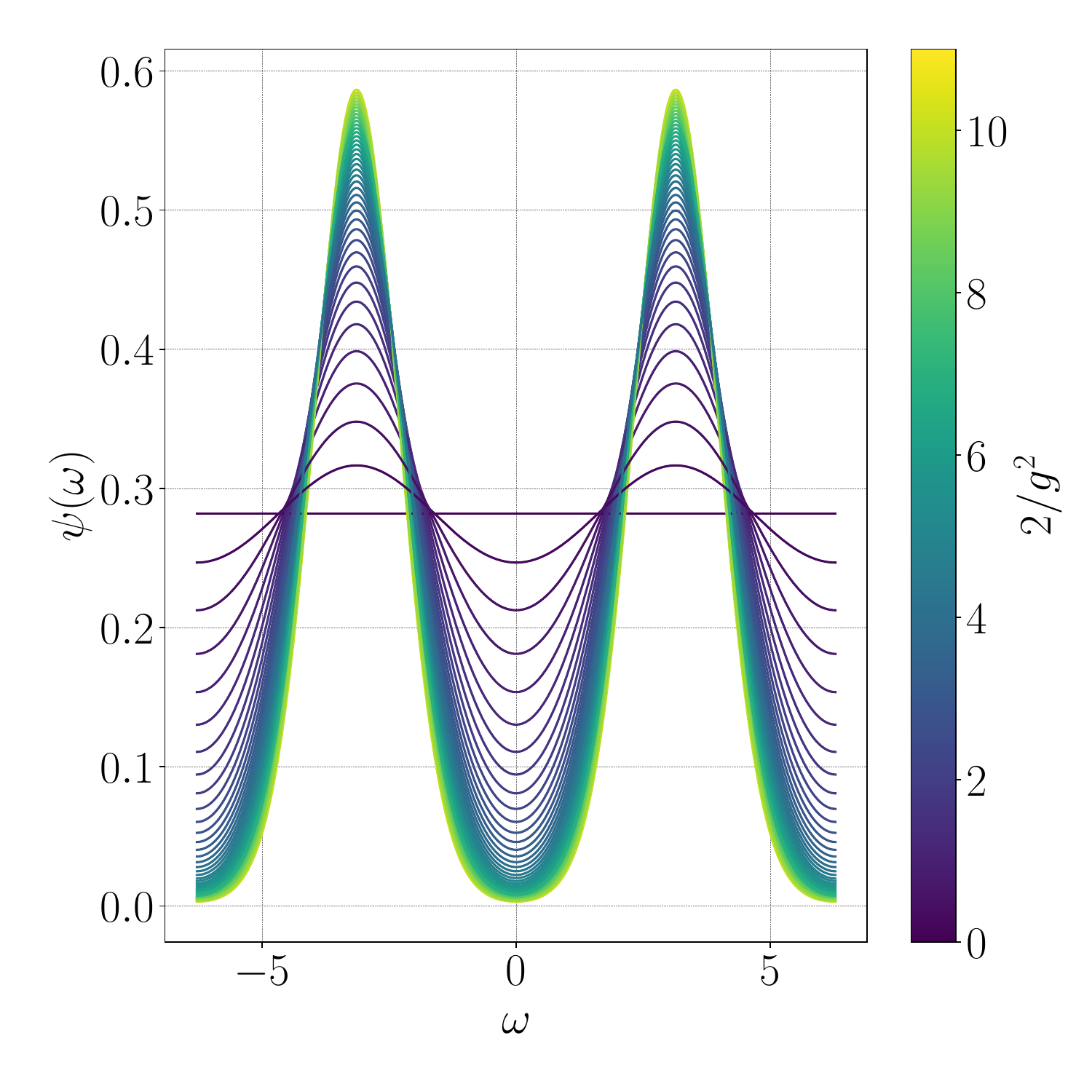}
    \includegraphics[width=0.5\textwidth]{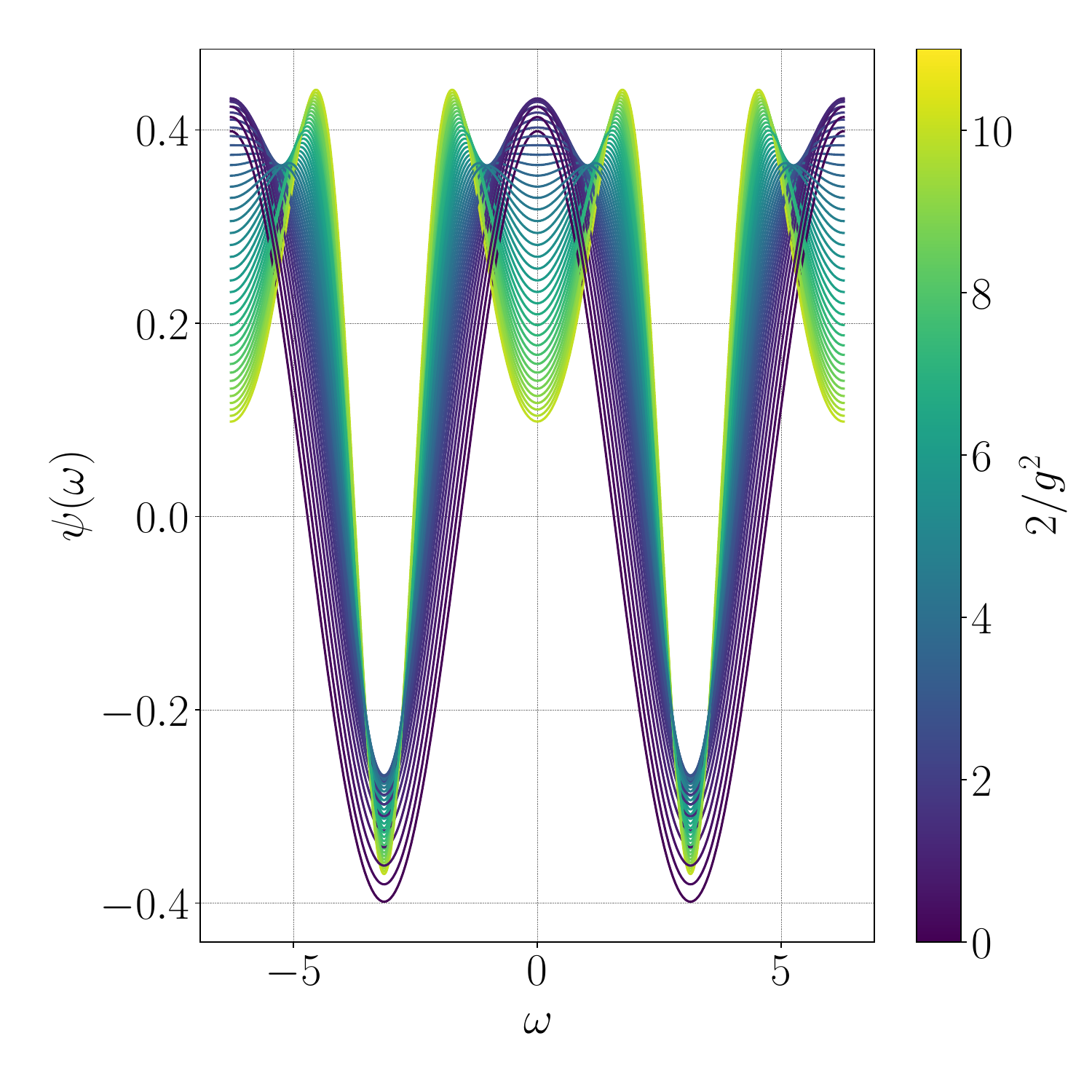}
    \caption{
        Learned eigenfunction $\psi$ for the ground state (left panel) and second excited state (right panel) of a single-plaquette $\mathrm{U}(1)$ system.
        On the horizontal axis, $\omega$ is such that $\cos{\omega}$ is the real part of the trace of the plaquette.
        Different curves correspond to different couplings, according to the colormap on the side. 
        Each curve is obtained by fine-tuning the model from the previous coupling. At ${1/g \to 0}$ the network is trained on the analytic expression.}
    \label{fig:U1EigenfunctionsTraining}
\end{figure}

\begin{figure}[H]
    \includegraphics[width=0.5\textwidth]{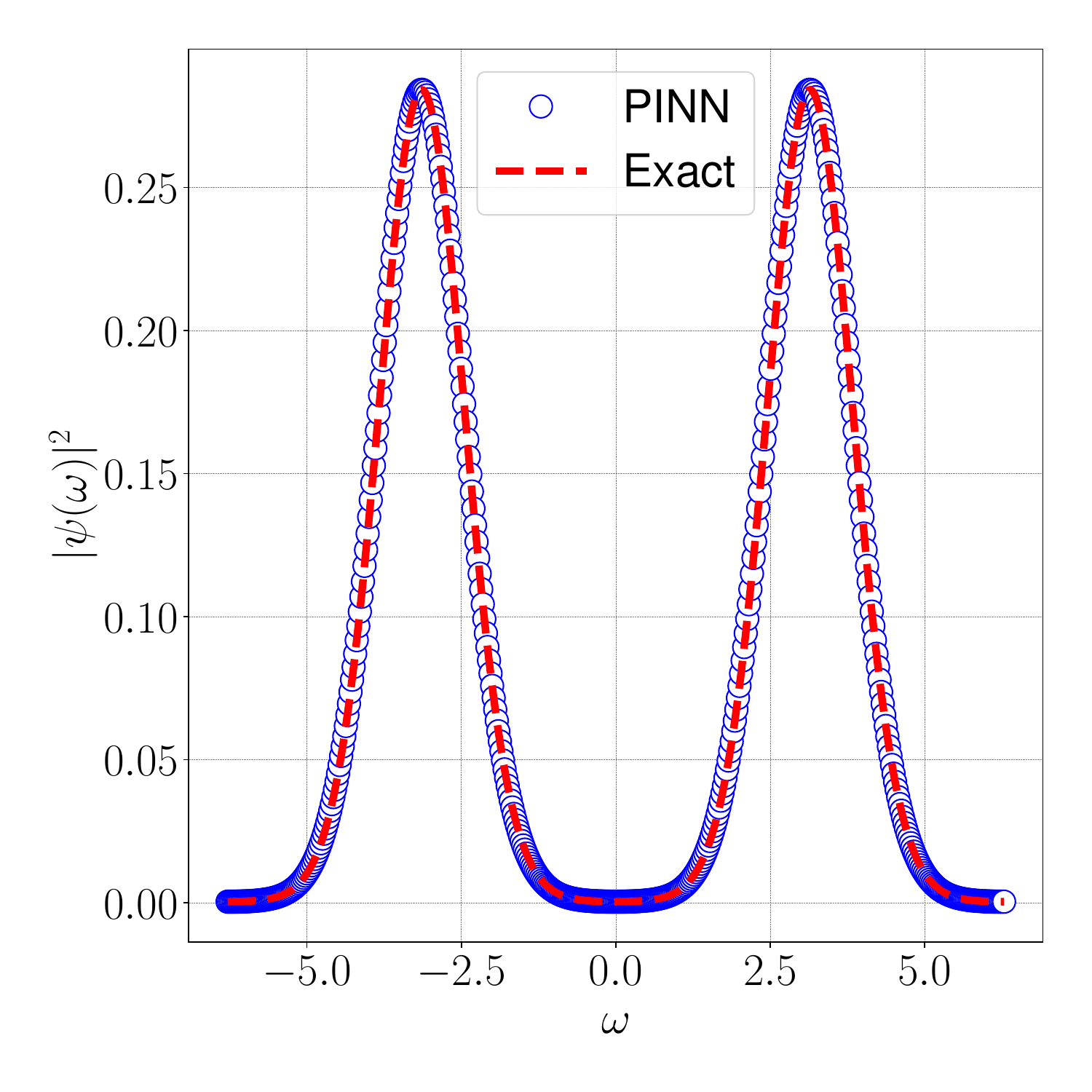}
    \includegraphics[width=0.5\textwidth]{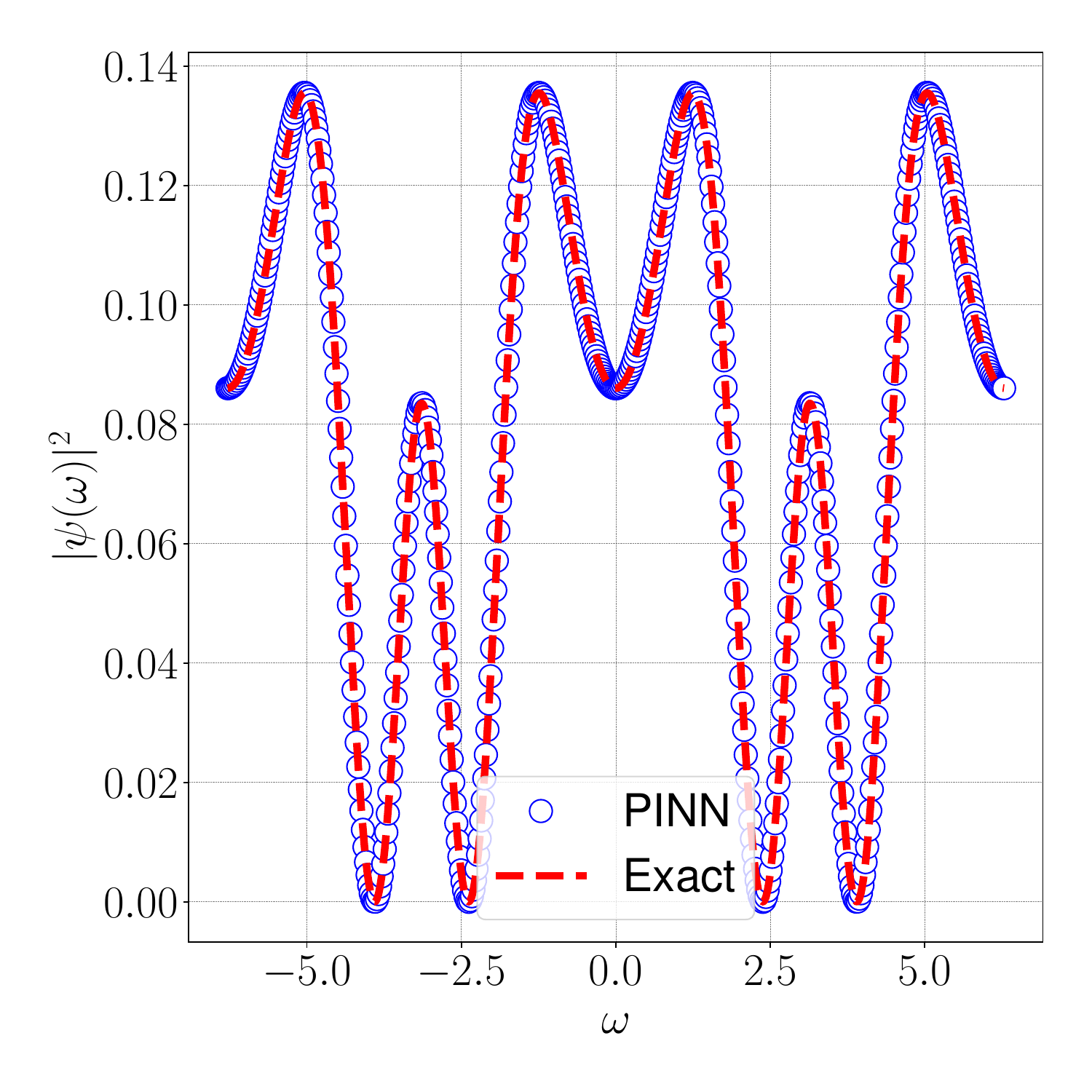}
    \caption{
        Comparison of learned eigenfunction (open points) and exact result (dashed line) at the intermediate coupling $2/g^2 = 5.0$.
        The left and right panel correspond respectively to the ground state and 2nd excited state eigenfunctions, for a $\mathrm{U}(1)$ single-plaquette system.}
    \label{fig:U1CompareEigenfunction}
\end{figure}

\begin{figure}[H]
    \includegraphics[width=0.5\textwidth]{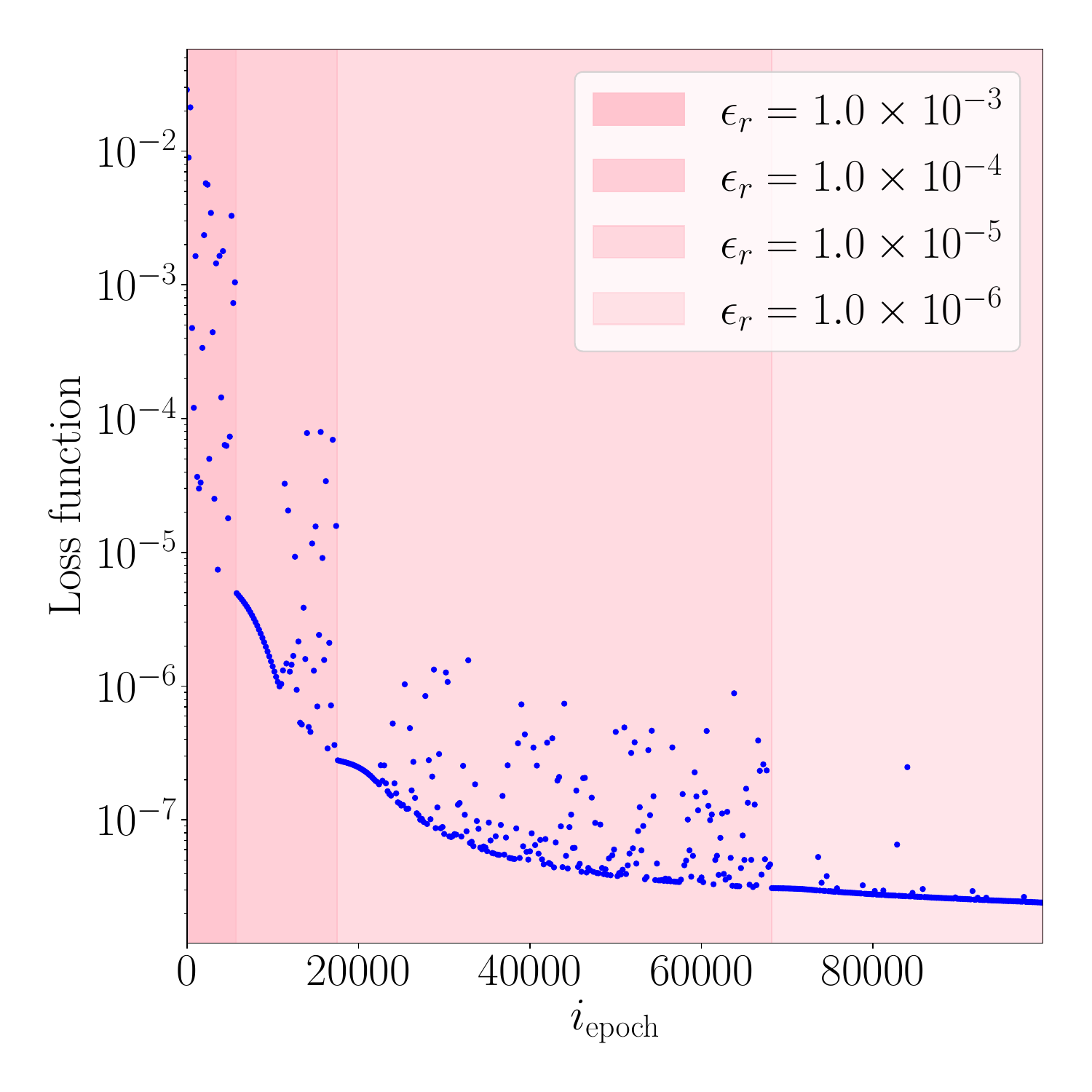}
    \includegraphics[width=0.5\textwidth]{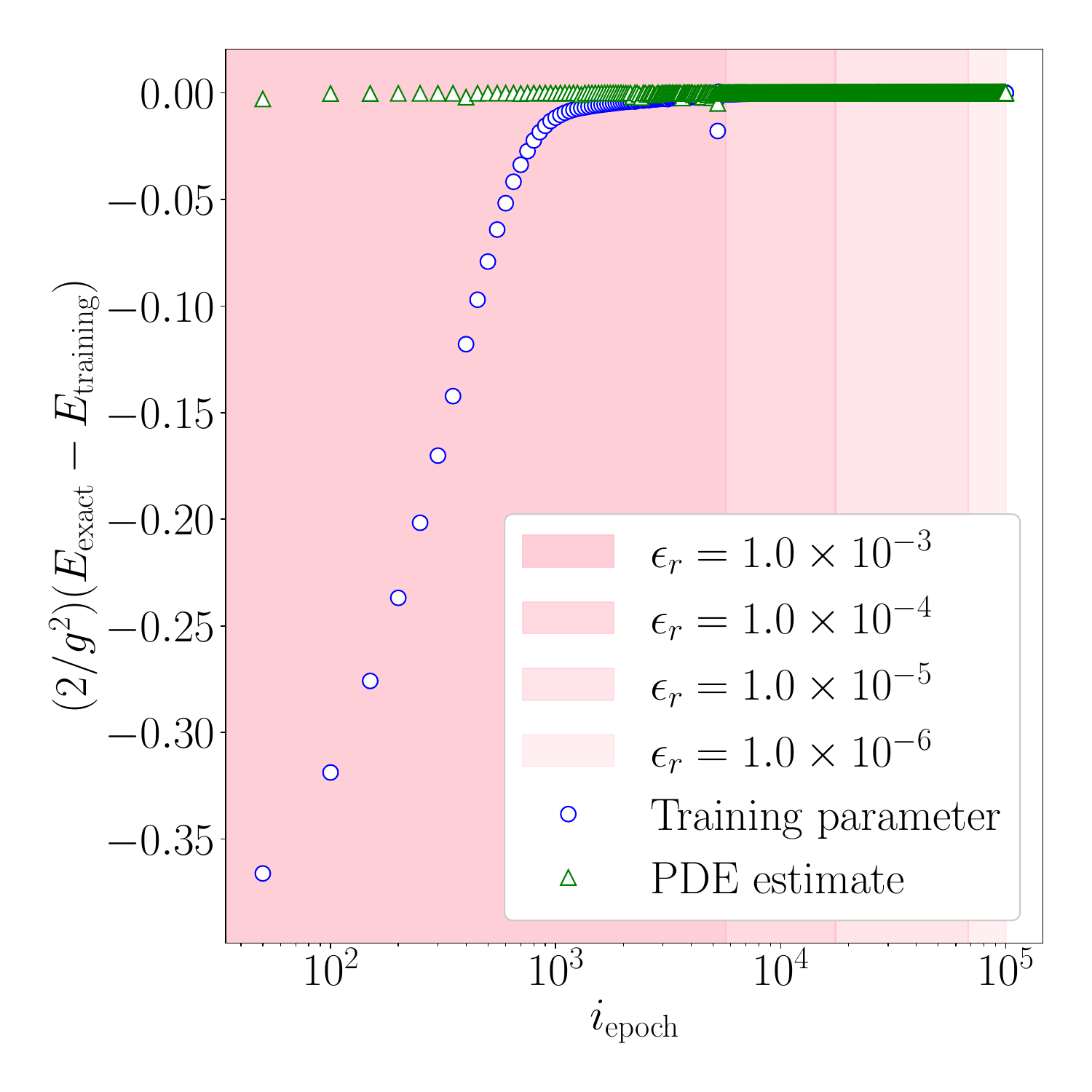}    
    \caption{
        Loss function (left) and energy (right) during the training, for the ground state of a single-plaquette $\mathrm{U}(1)$ system.
        The different color bands correspond to different learning rates $\epsilon_r$, scheduled according to a reduction-on-plateau criterion.
        The right panel shows the convergence of the energy parameter and the estimator of Eq.~\eqref{eq:EnergyEstimator} to the exact value.}
    \label{fig:U1TrainingGroundState}
\end{figure}

As described in Sec.~\ref{sec:PINNsStrategy}, for each eigenstate we first train a model on its strong coupling limit ($q=0$). From that model, we move to the weak-coupling
regime (larger $q$) by imposing the normalization condition, the
fulfillment of Eq.~\eqref{eq:MathieuEquationUone}, and the correct periodicity and
parity of the eigenfunction. 
In Fig.~\ref{fig:U1Energies} we show the energies of the first few eigenstates
determined with our PINN method, compared to the analytic results. 
In Fig.~\ref{fig:U1EigenfunctionsTraining} we show the learned wavefunction's coupling flow, 
and in Fig~\ref{fig:U1CompareEigenfunction} we plot the matching of the exact and trained eigenfunction. 
Finally, in Fig.~\ref{fig:U1TrainingGroundState} we show the typical behavior of the loss function and of the energy during the training.


\subsection{$\mathrm{SU}(2)$ gauge theory}

For $\mathrm{SU}(2)$ we have 3 generators in the Lie algebra, so that each of
the $N_\ell$ links is parametrized by 3 angles. 
A possible choice of the basis for the $k$-th link is~\cite{DAndrea:2023qnr}:
\begin{equation}
    \mathcal{U}_k = \mathcal{U}(\omega_k, \theta_k, \phi_k) =
    \begin{pmatrix}
        \cos{\frac{\omega_k}{2}} - i \sin{\frac{\omega_k}{2}} \cos{\theta_k} & -i \sin{\frac{\omega_k}{2}} \sin{\theta_k e^{-i \phi_k}}      \\
        -i \sin{\frac{\omega_k}{2}} \sin{\theta_k e^{i \phi_k}}        & \cos{\frac{\omega_k}{2}} + i\sin{\frac{\omega_k}{2}} \cos{\theta_k}
    \end{pmatrix}
    \, .
\end{equation}

In this basis the Hamiltonian reads~(see Sec.~\ref{sec:HamiltonianLGT}):
\begin{equation}
    H = \frac{g^2}{2} \sum_{k=1}^{N_\ell}
    \left[
    \frac{\mathbb{L}^2(\theta_k, \phi_k)}{4 \sin^2{\frac{\omega_k}{2}}} - \frac{\partial^2}{\partial \omega_k^2} - \cot{\frac{\omega_k}{2}} \frac{\partial}{\partial \omega_k}
    \right]
    - \frac{2}{g^2} \sum_{\Box} \cos{\varphi_\Box}
    \, ,
\end{equation}
where:
\begin{equation}
    \mathbb{L}^2(\theta_k, \phi_k) = -\frac{\partial^2}{\partial \theta_k^2} - \cot{\theta_k} \frac{\partial}{\partial \theta_k} - \frac{1}{\sin^2{\theta_k}} \frac{\partial^2}{\partial \phi_k^2}
    \, ,
\end{equation}
and $\cos{\varphi_\Box}$ is the real part of the trace of the product of links
around a plaquette $\Box$. %

When considering a single plaquette in $\mathrm{SU}(2)$, we have only one active link (see Sec.~\ref{sec:GaugeFixing}).
If we call $\omega$, $\theta$, $\phi$  the angles parametrizing it, the Hamiltonian becomes~(c.f.~Eq.\eqref{eq:HamiltonianMaxTreeSinglePlaquette}):
\begin{equation}
    H = 2 g^2 
    \left[
        \frac{\mathbb{L}^2(\theta, \phi)}{4 \sin^2{\frac{\omega}{2}}} - \frac{\partial^2}{\partial \omega^2} - \cot{\frac{\omega}{2}} \frac{\partial}{\partial \omega} 
    \right] 
    - \frac{4}{g^2} \cos{\frac{\omega}{2}}
    \, .
\end{equation}
We realize that $\mathbb{L}^2$ is the square of the usual angular momentum
operators~\cite{sakurai_napolitano_2017}, whose eigenfunctions are the
spherical harmonics $Y^{m}_{\ell}(\theta, \phi)$. Moreover, for this system,
the Gauss's law generators are the angular momentum operators in the angles
$\theta$ and $\phi$~\cite{DAndrea:2023qnr}. Thus, Gauss' law restricts the physical
subspace to the states with $\ell=m=0$, and all gauge-invariant wavefunctions
have the form:
\begin{equation}
    \psi(\omega, \theta, \phi) = Y^{0}_{0}(\theta, \phi) \frac{u(\omega)}{\sin{\omega/2}}
\end{equation}
In this way the time-independent Schr\"odinger equation becomes separable in
$\omega$, with a trivial dependence on $\theta$,$\phi$ fixed by gauge
invariance. The latter is a special feature of $\mathrm{SU}(2)$ and this choice
of angular basis.

The $\omega$-dependent part of the Hamiltonian on the physical space reads:
\begin{equation}
    H =
    2g^2
    \left(
    - \frac{d^2}{d \omega^2}
    - \cot{\frac{\omega}{2}} \frac{d}{d \omega}
    \right)
    - \frac{4}{g^2}
    \cos{\frac{\omega}{2}}
    \, .
\end{equation}
When acting $\frac{u(\omega)}{\sin{\omega/2}}$ the time-independent
Schr\"odinger equation ${H \psi = E \psi}$ simplifies to:
\begin{equation}
    \label{eq:SU2EigEquationInOmega}
    \left[
        -\frac{d^2}{d \omega^2}
        - \frac{1}{4}
        - \frac{2}{g^4} \cos{\frac{\omega}{2}}
        \right]
    u(\omega) = \frac{E}{2g^2} u(\omega)
    \, .
\end{equation}
In the variable $z=\omega/4$, Eq.~\eqref{eq:SU2EigEquationInOmega} becomes the Mathieu
equation~\cite{Ligterink:2000ug,Bender:2020jgr}~(similarly to the
$\mathrm{U}(1)$ case of Sec.~\ref{sec:UoneSinglePlaquetteResults}):
\begin{equation}
    \label{eq:MathieuEquationSU2}
    \left[ \frac{d^2}{d z^2} + \alpha - 2 q \cos{(2z)} \right] u(4z) = 0
    \, ,
\end{equation}
where ${\alpha=(4+8E/g^2)}$ and ${q=-16/g^4}$. 
In this case, the physical solutions $u(\omega)$ are the subset of Mathieu functions that are odd and $\pi$-periodic in $z$. 
In fact, the gauge invariant states are generated by traces of loops
(and their powers). In $\mathrm{SU}(2)$, the latter depend on $\omega$ only
through $\cos{\omega/2}$, which is even. Thus, $\psi(\omega) \sim
    \frac{u(\omega)}{\sin(\omega/2)}$ is even too, and $4\pi$-periodic in $\omega$
by construction.

\begin{figure}[H]
    \begin{center}
        \includegraphics[width=0.85\textwidth]{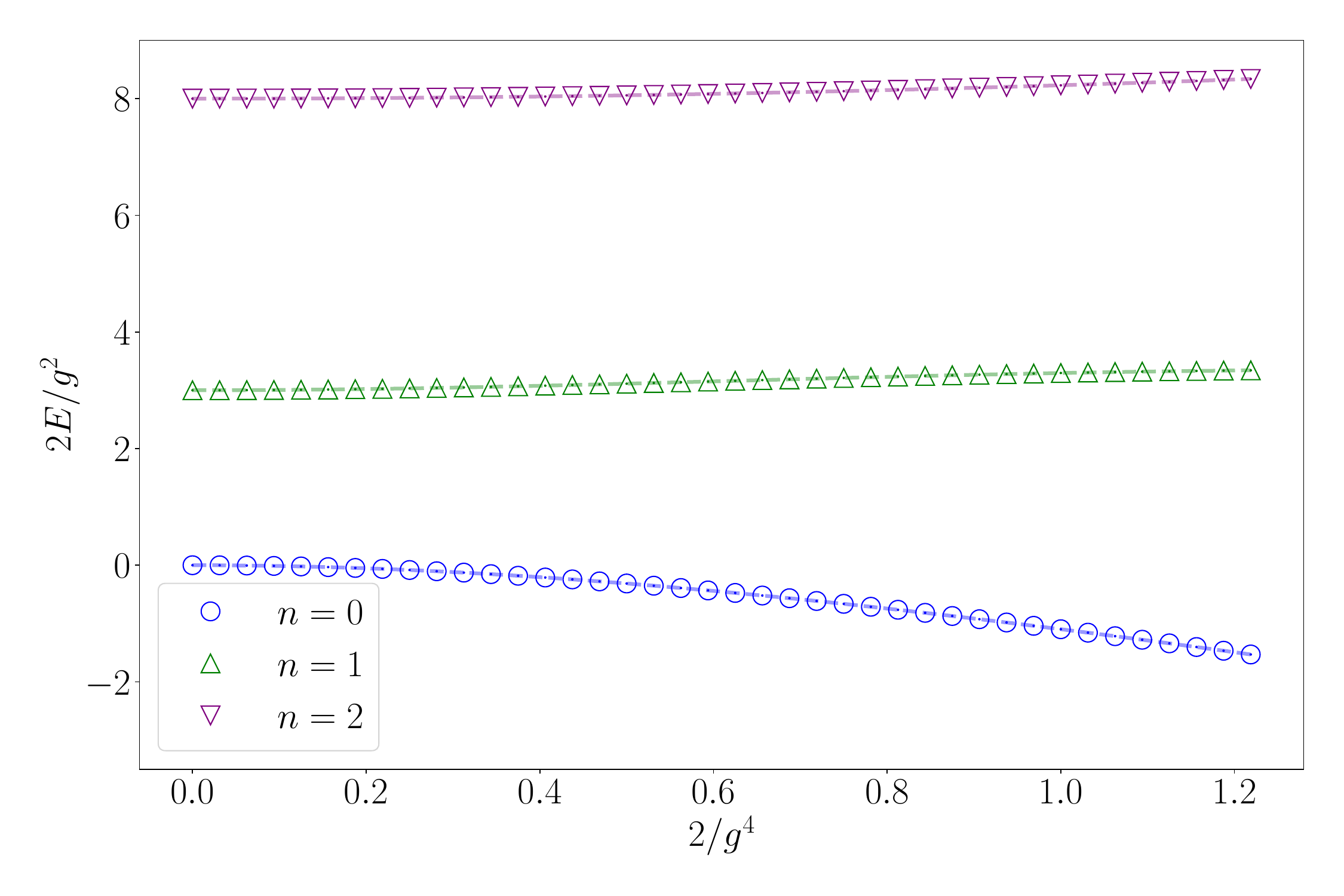}
    \end{center}
    \caption{
        Energy levels for a $\mathrm{SU}(2)$ single-plaquette system.
        The discrete points represent our prediction obtained with a PINN, while the dashed lines correspond to the exact values.}
    \label{fig:SU2Energies}
\end{figure}

\begin{figure}[H]
    \includegraphics[width=0.5\textwidth]{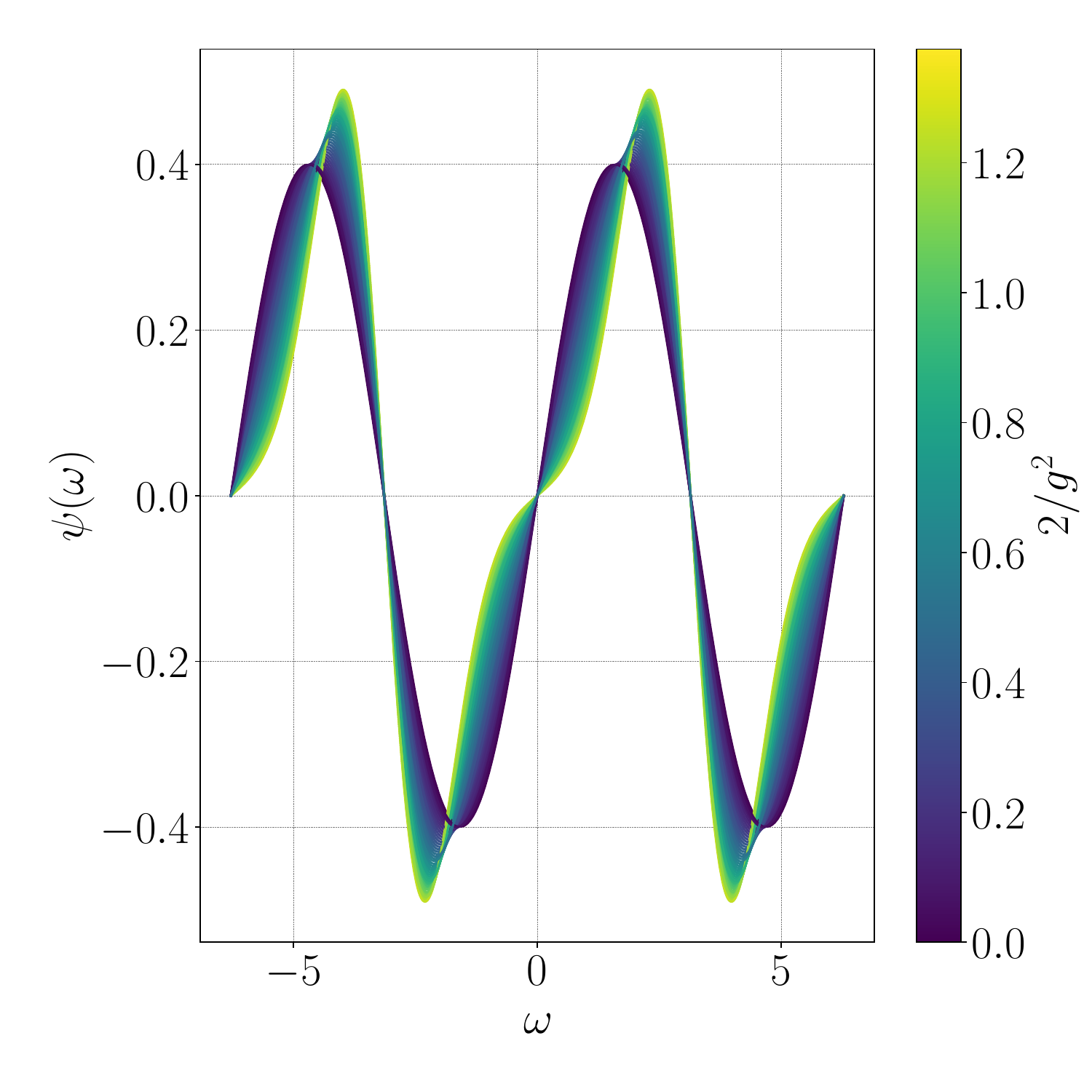}
    \includegraphics[width=0.5\textwidth]{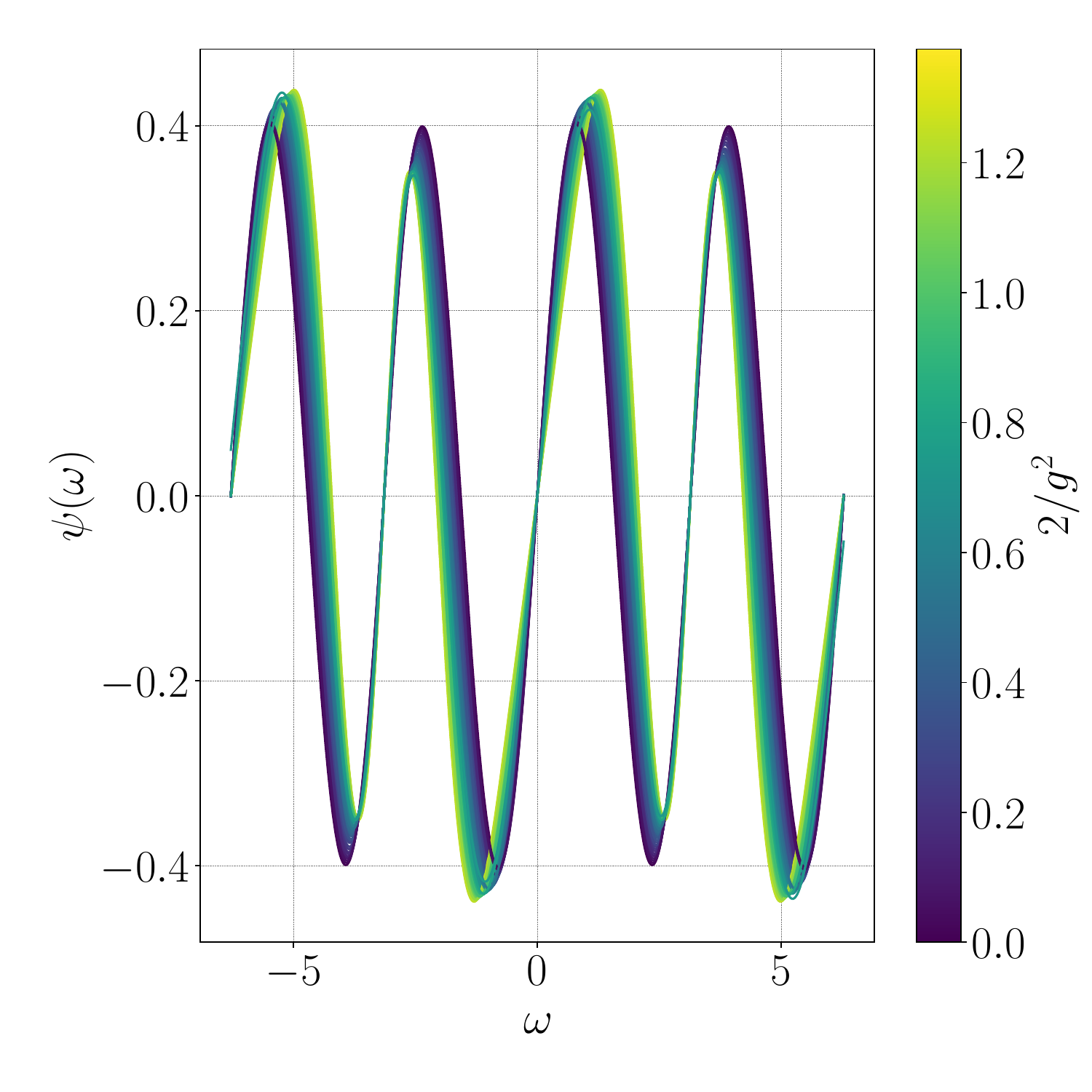}    
    \caption{
        Learned eigenfunction $\psi$ for the ground state (left panel) and first excited state (right panel) of a single-plaquette $\mathrm{SU}(2)$ system.
        On the horizontal axis, $\omega$ is such that $2\cos(\omega/2)$ is the real part of the trace of the plaquette.
        Different curves correspond to different couplings, according to the colormap on the side.
        Each curve is obtained by fine-tuning the model from the previous coupling. In the limit $1/g \to 0$ the network is trained on the analytic expression.
    }
    \label{fig:SU2EigenfunctionsTraining}
\end{figure}

\begin{figure}[H]
    \includegraphics[width=0.5\textwidth]{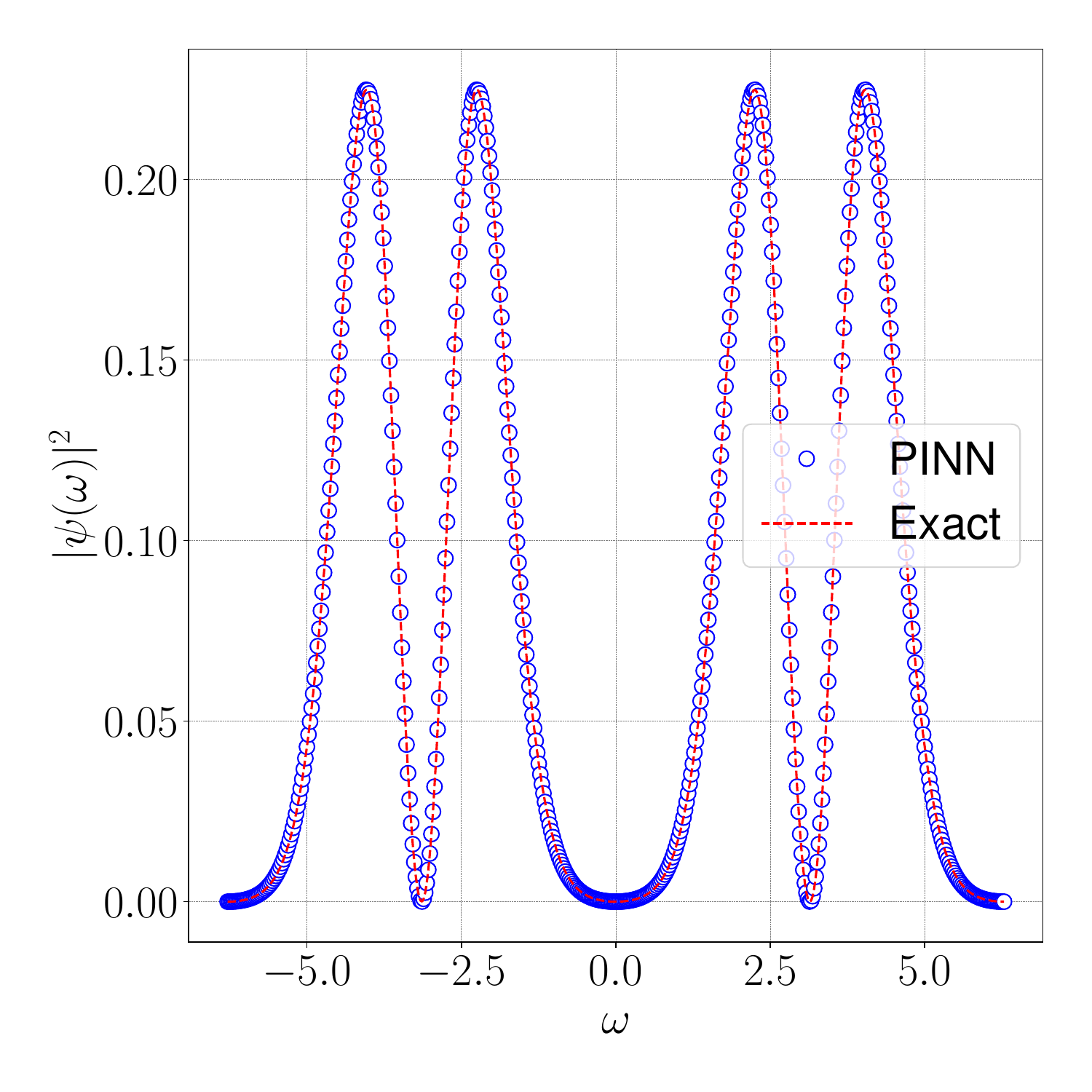}
    \includegraphics[width=0.5\textwidth]{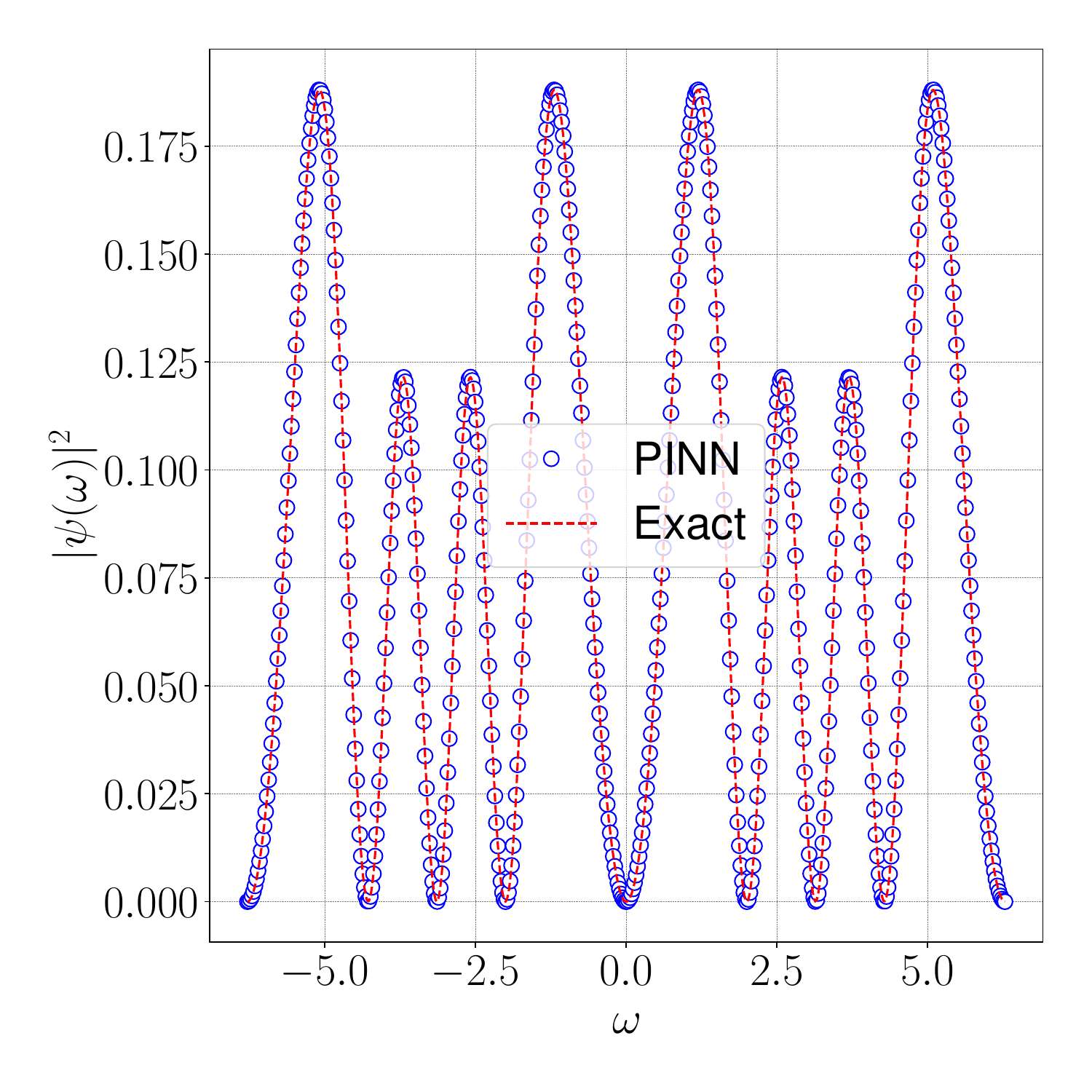}
    \caption{
        Comparison of learned eigenfunction (open points) and exact result (dashed line) at the intermediate coupling $2/g^2 = 1.0$.
        The left panel is the ground state and the right one the 2nd excited state eigenfunction.}
    \label{fig:SU2CompareEigenfunction}
\end{figure}

\begin{figure}[H]
    \includegraphics[width=0.5\textwidth]{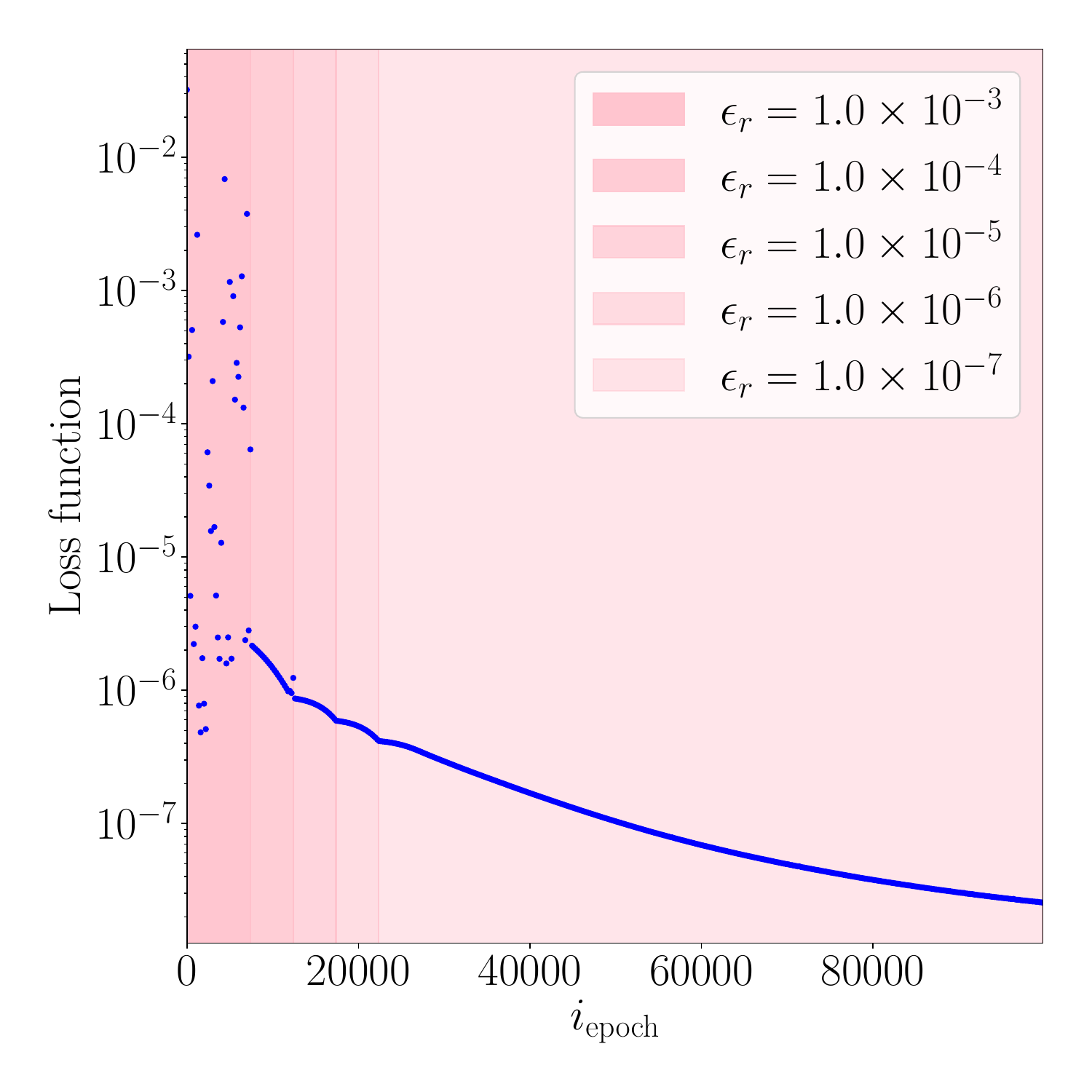}
    \includegraphics[width=0.5\textwidth]{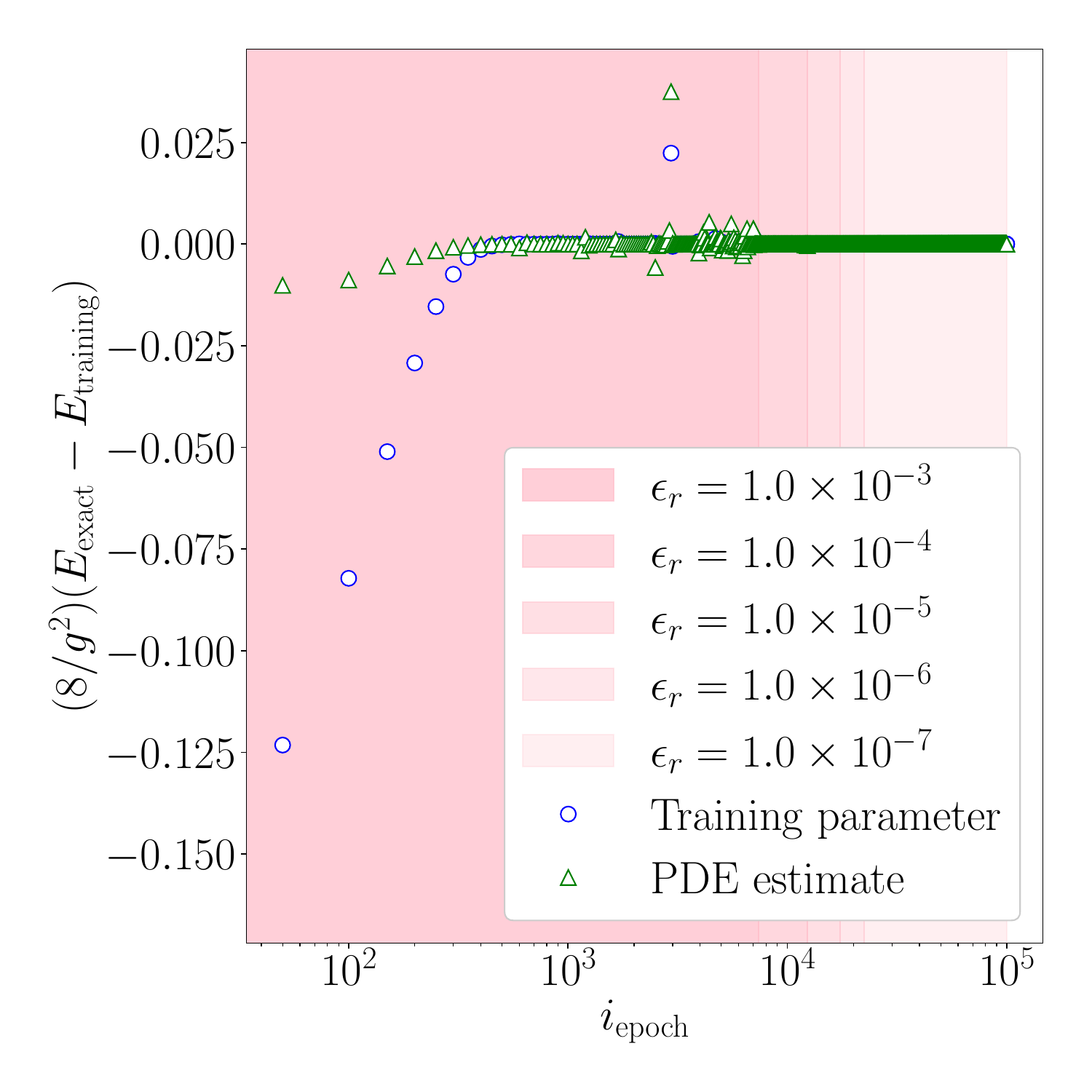}    
    \caption{
        \textit{Left panel}: Loss function during the training, for the ground state of a single-plaquette $\mathrm{SU}(2)$ system.
        \textit{Right panel}: convergence of the energy parameter and the estimator of Eq.~\eqref{eq:EnergyEstimator} with respect to the exact value.
        The different color bands correspond to different learning rates $\epsilon_r$, scheduled according to a reduction-on-plateau criterion.
        }
    \label{fig:SU2TrainingGroundState}
\end{figure}

As described in Sec.~\ref{sec:PINNsStrategy} (and as we did for $\mathrm{U}(1)$ in Sec.~\eqref{sec:UoneSinglePlaquetteResults}), we move to larger $q$ by imposing a unit normalization, the fulfillment of Eq.~\eqref{eq:MathieuEquationSU2}, the
correct periodicity, and the correct symmetry of the eigenfunction under a parity transformation. 
In Fig.~\ref{fig:SU2Energies} we show the learned energies of the first few eigenstates, as a function of the gauge coupling. 
In Fig.~\ref{fig:SU2EigenfunctionsTraining} we show the wavefunctions of the first few
eigenstates at different couplings, and in Fig.~\ref{fig:SU2CompareEigenfunction} we compare to the exact results. Finally,
in Fig.~\ref{fig:SU2TrainingGroundState} we show the typical
behavior of the loss function and of the energy during the training.


\section{Conclusions and outlook}
\label{sec:Conclusions}
In this work, we have introduced a neural network framework for solving $\mathrm{SU}(N_c)$ lattice gauge theories, verifying its applicability in the case of the single-plaquette spectrum of $\mathrm{U}(1)$ and $\mathrm{SU}(2)$ pure-gauge theories.

By incorporating the relevant physical constraints directly into Physics-Informed Neural Networks (PINNs), we have shown that one can reproduce the gauge coupling dependence of eigenfunctions and eigenvalues starting from the strong-coupling limit.
This is achieved with a novel method based on an adiabatic flow of the Yang-Mills gauge coupling: from a known analytic eigenstate and its energy at ${1/g \to 0}$, it is possible to determine its dependence at larger values of ${1/g}$.
The training process works by tuning the network to reproduce the correct wavefunction and learns its corresponding eigenvalue.
At ${1/g \neq 0}$, this is achieved by enforcing physical constraints on the wavefunction, without the need of a validation set from the exact solution.

The input to the network is a set of sampled gauge configurations.
This feature makes our PINN method a candidate for a scalable, multigrid-style path to approximate eigenstates of lattice gauge theories: one can train ``coarse'' representations (in configuration space), and progressively refine the sampling to fine-tune the eigenfunctions models.
Finally, the proposed adiabatic PINNs method provides a transparent avenue to the calculation of matrix elements from the definition, 

Future developments will focus on extending this framework to theories with fermionic degrees of freedom, to larger lattice volumes, and to gauge groups with higher color number $N_c$.

\section*{Acknowledgements}
We thank Prof.~Uwe-Jens Wiese for the very interesting and fruitful discussions that contributed to the development of this project.

This work was supported by the Swiss National Science Foundation (SNSF) through the grants \href{https://data.snf.ch/grants/grant/208222}{208222} and \href{https://data.snf.ch/grants/grant/10003675}{10003675}.
Calculations were performed on UBELIX \href{https://www.id.unibe.ch/hpc}{https://www.id.unibe.ch/hpc}, the HPC cluster at the University of Bern.

\appendix

\section{Hamitonian of $\mathrm{SU}(N_c)$ gauge theories on the lattice}
\label{sec:HamiltonianLGT}
The lattice action of an $\mathrm{SU}(N_c)$ gauge theory
reads~\cite{gattringer2009quantum}:
\begin{equation}
    S = S_\text{YM}[\mathcal{U}] + S_\text{F}[\psi, \bar{\psi}, \mathcal{U}]
    \, ,
\end{equation}
where the gauge links ${\mathcal{U}_\mu(x) \in \mathrm{SU}(N_c)}$ and the
fermionic fields ${\psi(x), \bar{\psi}(x)}$ are the degrees of freedom of the
theory. The Yang-Mills action $S_\text{YM}$ describes the dynamics of the
bosonic fields and their self-interactions, while the fermionic action
$S_\text{F}$ describes fermions and their interactions with the bosons. There
exist multiple regularizations for the lattice actions that differ only at
non-vanishing lattice spacing. In this work we limit our attention to
pure-gauge Yang-Mills (YM) theories, in the
standard Wilson
regularization~\cite{PhysRevD.15.1128,Romiti:2023hbd}:
\begin{equation}
    S_\text{YM}^\text{Wilson}
    = -\frac{2}{g^2} \sum_x \sum_{\mu = 0, \nu < \mu}^{d-1} \text{Re} \,
    \text{Tr}[\mathcal{U}_{\mu\nu}(x)]
    \, .
\end{equation}
The above sum runs over all spacetime points $x$, $d$ is the total number of
dimensions and:
\begin{equation}
    \mathcal{U}_{\mu\nu}(x) = \mathcal{U}_\mu(x) \mathcal{U}_\nu(x + \mu)
    \mathcal{U}_\mu^\dagger(x + \nu) \mathcal{U}_\nu^\dagger(x)
    \, .
\end{equation}

The invariance under $\mathrm{SU}(N_c)$ transformations allows to fix a gauge.
In the Weyl gauge ($\mathcal{U}_0(x) = \mathds{1}$), using the transfer matrix
formalism on the quantum theory and in the limit of vanishing temporal lattice
spacing, we find the Kogut-Susskind Hamiltonian~\cite{PhysRevD.15.1128,PhysRevD.20.2610,PhysRevD.15.1111,Romiti:2023hbd}~\footnote{
    This expression is consistent with,~e.g.,~Refs.~\cite{PhysRevD.15.1128,PhysRevD.20.2610,PhysRevD.15.1111,Romiti:2023hbd}.
    In the literature one can also find an additional factor $(1/2)$ in front of $H_B$
    (see~e.g.~Refs.~\cite{Ligterink:2000ug,Bender:2020jgr,DAndrea:2023qnr}). 
    The reason is that the choice of the constant $K$ in front of $H_B$ is just a convention in the definition of $g$ and of the lattice spacing $a$~\cite{PhysRevD.92.125018}.
    Two different choices of $K$ are related by matching the eigenvalues of the Hamiltonian in physical units: ${({K_2}/{K_1}) = \left({a_2}/{a_1}\right)^2 = \left({g_2}/{g_1}\right)^4}$.
    }:
\begin{equation}
    \label{eq:HamiltonianHBplusHE}
    H =
    H_{E} + H_{B} =
    \frac{g^2}{2} \sum_{\vec{x}} \sum_{\mu=1}^{d-1} \sum_{a=1}^{N_g}
    (L_a)_{\mu}^2(\vec{x})
    - \frac{1}{g^2} \sum_{\vec{x}} \sum_{\mu=1, \nu < \mu}^{d-1}
    \text{Tr}[U_{\mu\nu}(\vec{x}) + U_{\mu\nu}^\dagger(\vec{x})]
    \, .
\end{equation}
$H_E$ and $H_B$ are the \textit{electric} and \textit{magnetic} contributions to the Hamiltonian.
The sum in Eq.~\eqref{eq:HamiltonianHBplusHE} extends over the spatial lattice points and dimensions,
and $U_\mu(\vec{x})$ and $(L_a)_\mu(\vec{x})$ are now operators in the Hilbert
space $\mathcal{H}$. $N_g$ is the number of generators of the gauge group's Lie
algebra: ${N_g=1}$ for $\mathrm{U}(1)$ and ${N_g=N_c^2-1}$ for
$\mathrm{SU}(N_c)$. The \textit{magnetic} basis for $\mathcal{H}$ is given by
the configurations of links:
\begin{equation}
    \ket{\mathcal{U}} = \bigotimes_{\vec{x}} \bigotimes_{\mu=1}^{d-1}
    \ket{\mathcal{U}_{\mu}(\vec{x})}
    \, ,
    \quad
    U_{\mu}(\vec{x}) \ket{\mathcal{U}} = \mathcal{U}_{\mu}(\vec{x})
    \ket{\mathcal{U}}
    \, .
\end{equation}
The canonical momenta, for each $x$ and $\mu$, are the left and right
generators $L_a$, $R_a$. 
They satisfy the following properties and canonical
commutation relations with their conjugated link $U$~\cite{Romiti:2023hbd}:
\begin{align}
    [L_a, R_b] = 0 \,  ,
     & \quad
     L_a^2 = R_a^2 \, ,
    \\
    [L_a, L_b] = i f_{abc} L_c \,  ,
     & \quad
    [R_a, R_b] = i f_{abc} R_c \, ,
    \\
    [L_a, U] = -\tau_a U \, ,
     & \quad
    [R_a, U] = U \tau_a \, ,
\end{align}
where $f_{abc}$ are the structure constants of $\mathrm{su}(N_c)$ and $\tau_a$
its generators in color space. The residual gauge symmetry is generated by the
Gauss' law operator, so that the physical states are only the subset of
$\mathcal{H}$ satisfying:
\begin{equation}
    G_a(\vec{x}) |\psi\rangle_{\text{phys.}} = \left[ \sum_{\mu=1}^{d}
        (L_a)_{\mu}(\vec{x}) + (R_a)_{\mu}(\vec{x}-\mu) \right]
    |\psi\rangle_{\text{phys.}} = 0 \, .
\end{equation}

In the wavefunction formalism, in the links basis, the state of a system is
described by a functional ${\psi(\mathcal{U}) \in \mathbb{C}}$. The operators
$U_\mu(\vec{x})$ are represented as $\mathrm{SU}(N_c)$ matrices, and the $L_a,
    R_a$ by Lie derivatives~\cite{Jakobs:2023lpp,Romiti:2023hbd}:
\begin{align}
    (L_a)_\mu(\vec{x}) \, \psi(\ldots, \mathcal{U}_\mu(\vec{x}), \ldots) &= 
    - i \frac{d}{d \omega}
    \psi(\ldots, e^{-i \omega \tau_a} \mathcal{U}_\mu(\vec{x}), \ldots)
    \Big|_{\omega=0}
    \,	
    ,
    \\
    (R_a)_\mu(\vec{x}) \, \psi(\ldots, \mathcal{U}_\mu(\vec{x}), \ldots)
    &= 
    -i \frac{d}{d \omega}
    \psi(\ldots, \mathcal{U}_\mu(\vec{x}) e^{i \omega \tau_a} , \ldots)
    \Big|_{\omega=0}
    \,	
    .
\end{align}
The latter equation implies that
these are 1st-order differential operators. They can be also expressed in terms
of the variables $(\theta^a)_\mu(\vec{x})$ parametrizing the
$\mathcal{U}_\mu(\vec{x})$ as ${\mathcal{U}_\mu(\vec{x}) = \exp{(i
                (\theta^a)_\mu(\vec{x}) \, \tau_a)}}$. At given $N_c$, the coefficients in front of
each partial derivative $\partial_{\theta^a}$ can be found using the
Maurer-Cartan 1-form (see Ref.~\cite{Romiti:2023hbd} for a derivation).

\section{Gauge fixing}
\label{sec:GaugeFixing}

In this section we provide a concise recap on gauge fixing, focusing on the
peculiarities of the one-plaquette system discussed in
Sec.~\ref{sec:OnePlaquetteSystem}.

\begin{figure}[H]
    \centering
    \begin{tikzpicture}[scale=3.0]
        \filldraw[black] (0,0) circle (0.5pt);
        \filldraw[black] (1,0) circle (0.5pt);
        \filldraw[black] (1,1) circle (0.5pt);
        \filldraw[black] (0,1) circle (0.5pt);

        \filldraw[black] (2.5,0) circle (0.5pt);
        \filldraw[black] (3.5,0) circle (0.5pt);
        \filldraw[black] (3.5,1) circle (0.5pt);
        \filldraw[black] (2.5,1) circle (0.5pt);

        \draw[thick,dashed,->,-{Stealth[length=5pt,width=5pt]}] (0,0) -- (1,0) node[midway,below]
            {$\mathds{1}$};
        \draw[thick,dashed,->,-{Stealth[length=5pt,width=5pt]}] (1,0) -- (1,1) node[midway,right]
            {$\mathds{1}$};
        \draw[thick,->,-{Stealth[length=5pt,width=5pt]}] (0,1) -- (1,1) node[midway,above]
        {$\mathcal{P}={\mathcal{U}_1 \mathcal{U}_2 \mathcal{U}_3^\dagger
            \mathcal{U}_4^\dagger}$};
        \draw[thick,dashed,->,-{Stealth[length=5pt,width=5pt]}] (0,0) -- (0,1) node[midway,left] {$\mathds{1}$};

        \draw[thick,->,-{Stealth[length=5pt,width=5pt]}] (2.5,0) -- (3.5,0) node[midway,below]
            {$\mathcal{U}_4$};
        \draw[thick,->,-{Stealth[length=5pt,width=5pt]}] (3.5,0) -- (3.5,1) node[midway,right]
            {$\mathcal{U}_3$};
        \draw[thick,->,-{Stealth[length=5pt,width=5pt]}] (2.5,1) -- (3.5,1) node[midway,above]
            {$\mathcal{U}_2$};
        \draw[thick,->,-{Stealth[length=5pt,width=5pt]}] (2.5,0) -- (2.5,1) node[midway,left] {$\mathcal{U}_1$};

        \node[left] at (0,0) {$(0,0)$};
        \node[right] at (1,0) {$(1,0)$};
        \node[right] at (1,1) {$(1,1)$};
        \node[left] at (0,1) {$(0,1)$};

        \draw[thick,->] (1.3,0.5) -- (2.2,0.5)
        node[midway,above,yshift=2pt,align=center]
        {gauge \\ transformation};
    
        \node[left] at (2.5,0) {$(0,0)$};
        \node[right] at (3.5,0) {$(1,0)$};
        \node[right] at (3.5,1) {$(1,1)$};
        \node[left] at (2.5,1) {$(0,1)$};
    \end{tikzpicture}
    \caption{\textit{Left}: Maximal tree (dashed lines) and active link (solid
        line) for a single plaquette system with open boundary conditions.
        Every value of the active link can be gauged to an arbitrary
        configuration of the four links (\textit{right}), as in
        eq.~\eqref{eq:PlaqGaugeTransform}.}
    \label{fig:PlaquetteGaugeTransform}
\end{figure}
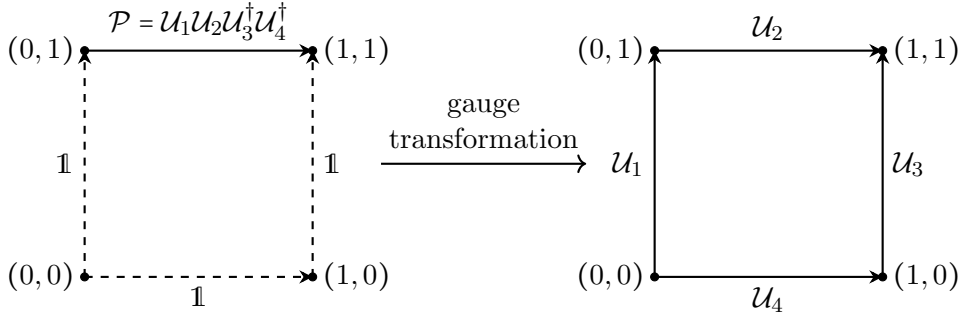

$\mathrm{SU}(N_c)$ gauge-symmetric field theories are
invariant under local group transformations~\cite{peskin2018introduction}. On
the lattice, the degrees of freedom transform as~\cite{gattringer2009quantum}:
\begin{align}
    \psi(x)              & \to \Omega(x) \psi(x)
                         & \text{(fermionic fields)} \, ,          \\
    \mathcal{U}_{\mu}(x) & \to \Omega(x)\, \mathcal{U}_{\mu}(x) \,
    \Omega^\dagger(x+a\hat{\mu})
                         & \text{(gauge links)} \, ,
\end{align}
where $\Omega(x)$ is a group element in the fundamental representation, $a$ is
the lattice spacing and $\hat{\mu}$ is the unit vector in the $\mu$-th axis.
As a consequence, the description of the system contains some redundancy: some
of the gauge links can be mapped to the identity without changing the content
of the theory. Using the symmetry to fix some of the links to $\mathds{1}$ is
called \textit{gauge fixing}, and the set with the largest number of
gauge-fixable links constitutes a \textit{maximal
    tree}~\cite{PhysRevD.15.1128,Ligterink:2000ug}. Geometrically, the latter
corresponds to a maximal set of links on the lattice that do not form any
closed loop.
For a single plaquette with open boundary conditions for instance, the
gauge-invariant combination ${\mathcal{P}={\mathcal{U}_1 \mathcal{U}_2
                    \mathcal{U}_3^\dagger \mathcal{U}_4^\dagger}}$ is the only \textit{active}
link, from which any link configuration is obtained by a gauge
transformation~\cite{DAndrea:2023qnr}:
\begin{equation}
    \label{eq:PlaqGaugeTransform}
    \ket{
        \mathds{1},
        \mathcal{U}_1 \mathcal{U}_2 \mathcal{U}_3^\dagger
        \mathcal{U}_4^\dagger,
        \mathds{1},
        \mathds{1}
    }
    \to
    \ket{
        \mathds{1},
        \mathcal{U}_1 \mathcal{U}_2,
        \mathcal{U}_4 \mathcal{U}_3,
        \mathds{1}
    }
    \to
    \ket{
        \mathds{1},
        \mathcal{U}_1 \mathcal{U}_2,
        \mathcal{U}_3,
        \mathcal{U}_4
    }
    \to
    \ket{\mathcal{U}_1,\mathcal{U}_2,\mathcal{U}_3,\mathcal{U}_4}
    \, ,
\end{equation}
where $1,2,3,4$ indicate the indices of the links in the plaquette as in
Fig.~\ref{fig:PlaquetteGaugeTransform}.

In the Hamiltonian formalism, gauge symmetry is a result of the residual gauge
freedom in the Weyl gauge, $U_0(x)=\mathds{1}$, at each
timeslice~\cite{PhysRevD.15.1128}. In practice, when solving the theory we can
either impose gauge-symmetry by hand through Gauss' law, work in a gauge-fixed
Hilbert space directly or, equivalently, formulate the Hamiltonian and the
Hilbert space in terms of gauge-invariant operators.

\begin{figure}[H]
    \begin{center}
        \begin{tikzpicture}[scale=1.5]
            \def\Nx{4}
            \def\Ny{4}
            \def\cx{2}
            \def\cy{2}

            \foreach \x in {0,...,4} {
                    \foreach \y in {0,...,4} {
                            \filldraw[black] (\x,\y) circle (1.2pt);
                        }
                }

            \foreach \x in {0,...,4} {
                    \foreach \y in {0,...,3} {
                            \draw[thick,dashed] (\x,\y) -- (\x,{int(\y+1)});
                        }
                }

            \foreach \y in {0,...,4} {
                    \foreach \x in {0,...,3} {
                            \ifnum\y=\cy
                                \draw[thick,dashed] (\x,\y) -- ({int(\x+1)},\y);
                            \else
                                \draw[thick] (\x,\y) -- ({int(\x+1)},\y);
                                \pgfmathtruncatemacro{\xshift}{\x-2}
                                \pgfmathtruncatemacro{\yshift}{\y-2}
                            \fi
                        }
                }

            \node[black,above] at (\cx+1.5,\cy+2.05){$\kappa$};

            \filldraw[blue] (\cx,\cy) circle (2.2pt);
            \node[below right,blue] at (\cx,\cy-0.075) {$(0,0)$};

            \draw[very thick,red,->,-{Stealth[length=5pt,width=5pt]}] (\cx,\cy+0.075) -- (\cx+1-0.075,\cy+0.075);
            \draw[very thick,red,->,-{Stealth[length=5pt,width=5pt]}] (\cx+1-0.075,\cy+0.075) --
            (\cx+1-0.075,\cy+2+0.075); 
            \draw[very thick,red,->,-{Stealth[length=5pt,width=5pt]}] (\cx+1-0.075,\cy+2+0.075) --
            (\cx+2+0.075,\cy+2+0.075); 
            \draw[very thick,red,->,-{Stealth[length=5pt,width=5pt]}] (\cx+2+0.075,\cy+2+0.075) --
            (\cx+2+0.075,\cy-0.075); 
            \draw[very thick,red,->,-{Stealth[length=5pt,width=5pt]}] (\cx+2+0.075,\cy-0.075) -- (\cx,\cy-0.075);

            \node[red,below right] at (\cx+2.1,\cy+2.0) {$\mathcal{P}(\kappa)$};

        \end{tikzpicture}
    \end{center}

    \caption{
        Construction of loop operators on a lattice with 2 spatial dimensions,
        analogous to Ref.~\cite{DAndrea:2023qnr}.
        The dashed lines correspond to the links gauged to $\mathds{1}$ in the
        maximal tree, while the solid lines are the active links.
        The loop is built walking from the origin and going through the
        respective active link.
        An explicit construction is shown for $P(\kappa)$,
        corresponding to an active link $\kappa$.
        The (untraced) loops are still gauge-variant under a gauge transformation at the
        origin.}
    \label{fig:MaxTree}
\end{figure}
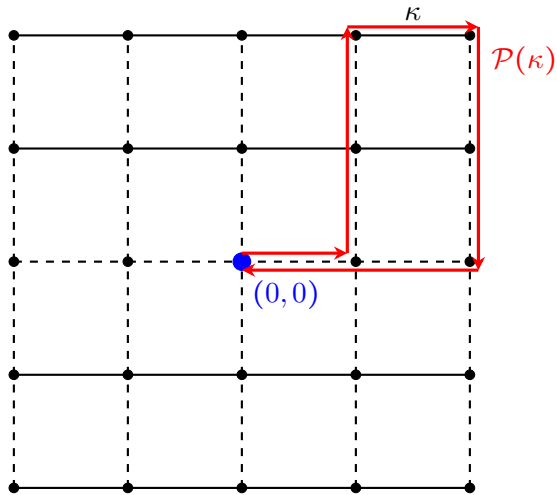

For the purpose of this work and in the context of pure-gauge
$\mathrm{SU}(N_c)$ gauge theories, we recall here the hybrid approach adopted
in Ref.~\cite{DAndrea:2023qnr}. In fact, it turns out that we can write the
Hamiltonian in terms of loop operators (and their canonical momenta) that are
gauge invariant up to a gauge transformation at the origin of the lattice.
Fig.~\ref{fig:MaxTree} provides a sketch of the loops construction in $2$
spatial dimensions.

In particular, for the case of the one-plaquette problem analyzed in
Sec.~\ref{sec:OnePlaquetteSystem}, the gauge-fixed Hamiltonian of
Eq.~\eqref{eq:HamiltonianHBplusHE} reads:
\begin{equation}
    \label{eq:HamiltonianMaxTreeSinglePlaquette}
    H = H_E + H_B =
    2 {g^2} L^2_a(\kappa) - \frac{2}{g^2}
    \operatorname{Re}\operatorname{Tr}{P(\kappa)}
    \, ,
\end{equation}
where $\kappa$ is the only active link of the one-plaquette system, $P(\kappa)$
the corresponding loop variable and $L_a(\kappa)$ are its associated canonical
momenta. In the magnetic basis, $L_a(\kappa)$ is a differential operator in the
agles defining $P(\kappa)$~\cite{Romiti:2023hbd}. If we use
${\omega=\operatorname{Arg}(\operatorname{Tr}{P(\kappa)})}$ as one of the
angles, $H_B$ will depend only on $\omega$. In particular, for $\mathrm{U}(1)$
and $\mathrm{SU}(2)$ there is only one Casimir operator, thus $H_E$ can depend
only on $\omega$ because of gauge invariance. As a result, the time-independent
Schr\"odinger equation for a one-plaquette system reduces to an ODE for the
eigenvalues. For $N_c>1$ the gauge is not fully fixed, and we also need to
impose the fulfillment of the PDE corresponding to Gauss' law at the
origin~\cite{Ligterink:2000ug,DAndrea:2023qnr}.

\printbibliography

\end{document}